\documentclass[aps,a4paper, preprint, superscriptaddress,preprintnumbers,floatfix,nofootinbib,amsmath,amssymb]{revtex4}
\usepackage{url}
\usepackage{hyperref}
\usepackage{cleveref}

\usepackage{amstext,amssymb}
\usepackage{amsmath}
\usepackage{graphicx}
\usepackage{soul}
\usepackage{xspace}
\usepackage{color}
\DeclareUnicodeCharacter{2212}{-}
\usepackage{units}
\usepackage{slashed} 
\usepackage{multirow}
\newcommand{\be}{\begin{equation}}
\newcommand{\ee}{\end{equation}}
\newcommand{\bea}{\begin{eqnarray}}
\newcommand{\eea}{\end{eqnarray}}
\newcommand{\nn}{\nonumber}
\usepackage{placeins}

\def\s1{\hat s}

\usepackage{slashed}

\newcommand{\nua}[1]{\ensuremath{\rlap{\kern-2.5pt\ensuremath{\overset{\scriptscriptstyle(-)}{\phantom{\nu}}}}{\ensuremath{{\nu}_{#1}}}}\xspace}

\begin{document}
\title{Implications of $A_4$ modular symmetry on neutrino mass, mixing and leptogenesis with linear seesaw}
\author{ Mitesh Kumar Behera}
\email{miteshbehera1304@gmail.com}
\affiliation{School of Physics,  University of Hyderabad, Hyderabad - 500046,  India}
\author{Subhasmita Mishra}
\email{subhasmita.mishra92@gmail.com}
\affiliation{Department of Physics, IIT Hyderabad,Kandi - 502285, India}
\affiliation{School of Applied Sciences, Centurion University of Technology and Management, Odisha - 761211}
\author{ Shivaramakrishna Singirala}
\email{krishnas542@gmail.com}
\affiliation{School of Physics,  University of Hyderabad, Hyderabad - 500046,  India}
\affiliation{Discipline of Physics, Indian Institute of Technology Indore, Simrol, Indore-453 552, India}
\author{ Rukmani Mohanta}
\email{rmsp@uohyd.ac.in}
\affiliation{School of Physics,  University of Hyderabad, Hyderabad - 500046,  India}

\begin{abstract}
The present work is inspired by the application of $A_4$ modular symmetry in the linear seesaw framework, which restricts  the use of multiple flavon fields. Linear seesaw is realized with six heavy $SU(2)_L$ singlet fermion superfields and a weighton in a supersymmetric framework. The non-trivial transformation of Yukawa couplings under the $A_4$ modular symmetry helps to explore the neutrino phenomenology with a specific flavor structure of the mass matrix. We discuss the phenomena of neutrino mixing and show that the obtained  mixing angles and CP violating phase in this framework are compatible with the observed $3\sigma$ range of the current oscillation data. In addition, we also  investigate the non-zero CP asymmetry from the decay of lightest heavy fermion superfield to explain the preferred phenomena of baryogenesis through leptogenesis including flavor effects.
\end{abstract}

\maketitle
\flushbottom

\section{INTRODUCTION}
\label{sec:intro}

Standard Model (SM) is not triumphant concering the observed properties of neutrinos, i.e. they are not exactly massless as predicted in the SM, but posses tiny but non-zero masses \cite{Faessler:2020sgs,Aker:2019uuj,Pontecorvo:1967fh,Feruglio:2017ieh}, as inferred from neutrino oscillation data. The phenomenon of neutrino oscillation is now well-established, which provides strong  evidence  for the  mixing of neutrinos and atleast two of them have non-zero masses \cite{Tanabashi:2018oca}.  It is well evident from theory and experiments that neutrinos don't have right-handed (RH) counterparts in the SM, which makes them unfavorable to have Dirac mass, like other charged fermions, nonetheless, dimension-five Weinberg operator \cite{Weinberg:1980bf, Weinberg:1979sa,Wilczek:1979hc} can be  useful for providing them masses. However,  the origin and flavour structure of this operator is under arguable terms.
 As a result, exploring scenarios beyond the standard model (BSM)  becomes crucial in generating non-zero masses for neutrinos. There exists numerous models in the literature to explain the observed  data from various neutrino oscillation experiments,  for example, the most popular seesaw mechanism \cite{Minkowski:1977sc, Mohapatra:1979ia, GellMann:1980vs},  radiative mass generation \cite{Zee:1980ai, Babu:1988ki}, extra-dimensions \cite{ArkaniHamed:1998vp}, etc. A prevalent feature of many BSM scenarios, which elucidate the generation of non-zero neutrino masses, is the existence of sterile neutrinos, which are   SM gauge singlets, generally considered as right-handed  neutrinos, coupled to the standard  active neutrinos through  Yukawa interactions. A priori, their masses and interaction strengths can span over many orders of magnitude, which thus lead to a wide variety of observable phenomena. For example, in the canonical  seesaw framework, to explain the eV-scale light neutrinos, the  RH neutrino mass is supposed to be of the order of $10^{15}$ GeV,  which is obviously beyond the reach of current as well as future experiments.
 However, its low scale variants like inverse seesaw \cite{Mohapatra:1986bd, GonzalezGarcia:1988rw,CarcamoHernandez:2019pmy} linear seesaw \cite{Malinsky:2005bi}, extended seesaw \cite{Mohapatra:2005wg}, etc., where the heavy neutrino mass can be in the TeV range, which makes them experimentally verifiable. \vspace{1mm}
 
On the other hand, the non-abelian discrete flavor symmetry group $A_4$ provides a possible underlying symmetry for the neutrino mass matrix \cite{Ma:2001dn}, which however yields a vanishing reactor mixing angle $\theta_{13}$.  Nonetheless,  it has still been widely used to describe the neutrino mixing phenomenology with inclusion of simple perturbation by introducing extra flavon fields, which are SM singlets but transform non-trivially under the flavor symmetry group,  leading to  nonzero reactor mixing angle. Thus, the flavons become  integral part in realizing the observed pattern in neutrino mixing due to their particular vacuum alignment,  which play a crucial role in spontaneous breaking of the discrete flavor symmetry \cite{Pakvasa:1977in}. 
Typically, flavons, in quite a number are necessary to realize certain phenomenological aspects under the framework of such flavor symmetry. However, there are additional drawbacks to this approach, where higher dimensional operators can ruin the predictability of the discrete flavor symmetry. Furthermore, the customary use of flavor symmetry is to constrain the  mixing angles while neutrino masses remain undetermined except in few scenarios. These drawbacks are eliminated by making a modular invariance approach \cite{Feruglio:2017spp}.\vspace{1mm}

Present, proposition of modular flavor symmetries has been carried out in the literature \cite{Feruglio:2017ieh,Feruglio:2017spp,King:2020qaj} to bring predictable flavor structures into limelight. Some of the effective models of modular symmetry that have recently been investigated \cite{King:2017guk,Altarelli:2010gt,Ishimori:2010au,King:2015aea}, do not make use of the flavon fields apart from the modulus $\tau$,  and  hence, the flavor symmetry   is broken when this complex modulus $\tau$ acquires vacuum expectation value (VEV). The usage of perplexed vacuum alignment is avoided, the only need is a mechanism which can fix the modulus $\tau$. As a result, this framework transforms Yukawa couplings, where these couplings are function of modular forms, which indeed are holomorphic function of $\tau$. To put it differently, these couplings transpire under a non-trivial representation of a non-Abelian discrete flavor symmetry approach,  such that they can compensate the use of flavon fields, which indeed are not required or minimized in realising the flavor structure. In the above context, after going through myriad texts, it was comprehended that there are many groups available e.g., the modular group of $A_4$ \cite{Abbas:2020qzc,King:2020qaj,Wang:2019xbo,Lu:2019vgm, Kobayashi:2019gtp,Nomura:2019xsb,Asaka:2019vev}, $S_4$ \cite{Penedo:2018nmg,Liu:2020akv,Gui-JunDing:2019wap,Kobayashi:2019xvz,Novichkov:2018ovf,Novichkov:2020eep}, $A_5$ \citep{Ding:2019zxk,Novichkov:2018nkm}, larger groups \cite{Kobayashi:2018wkl}, various other modular symmetries and double covering of $A_4$ \cite{Nomura:2019lnr,Ma:2015fpa,Mishra:2019oqq}, where prediction of masses, mixing, and CP phases related to quarks and/or leptons are investigated. \vspace{1mm}

It is worth realizing that neutrino mass models which are based on modular invariance could involve only few coupling  strengths so that neutrino masses and mixing parameters are correlated. However, there is an extension of above formalism to combine it with the generalized CP symmetry \cite{Acharya:1995ag,Lu:2019vgm,Novichkov:2019sqv,Baur:2019kwi,Dent:2001cc,Giedt:2002ns}. As we know that, $S$ and $T$ representation are symmetric, so the modular form multiplets, if normalized aptly, acquire complex conjugation under CP transformation. As a result, all the couplings get constrained due to generalized CP symmetry in a modular invariant model to be real \cite{Novichkov:2019sqv}, hence, the model prediction power gets meliorated. To implement the aforesaid, it is very intriguing to see the application of modular symmetry in establishing a model for neutrino mass generation as it would envisage for the signals of new physics through the observables in  neutrino sector \cite{Chen:2019ewa}.
\vspace{1mm}
 
In this paper, we intend to  examine the advantages of $A_4$ modular symmetry by applying it to linear seesaw mechanism in supersymmetric (SUSY) context. The linear seesaw  formalism requires three left-handed neutral fermions $S_{L_i}$ in addition to three-right handed ones $N_{R_i}$ $(i=1,2,3)$ and generates the neutrino mass matrix which is intricate enough, and has been studied in the context of $A_4$ symmetry in  \cite{Sruthilaya:2017mzt,Borah:2018nvu,Borah:2017dmk}. Furthermore, $S_{L_i}$ \& $N_{R_i}$ are assigned as triplets under $A_4$ symmetry and Yukawa couplings are expressed in  modular forms by which the neutrino mass matrix attains a constrained structure.  Consequently, numerical analysis is performed to scan for free parameters of the model and to look for the  region which can fit neutrino oscillation data. After obtaining the constraints on  the model parameters, neutrino sector observables are predicted.
It should be noted that the imposition of modular  symmetry rather simplifies the inclusion of multiple flavons (i.e.,  weighton in SUSY), which complicates the problem of vacuum alignments  in usual $A_4$ scenario.  However, apart from model building perspective,  essentially there are no distinct  phenomenological differences between the two scenarios which could distinguish them, as in both the approaches, the singlet fermions $N_{Ri}$ and $S_{Li}$ are required.
\vspace{1mm}

Structure of this paper is as follows. In Sec. \ref{sec:linear} we outline the well known linear seesaw mechanism with discrete $A_4$-modular flavor symmetry and its appealing features resulting in simple mass structure for the charged leptons and neutral leptons including light active neutrinos and other two types of sterile neutrinos. We then provide a discussion for the light neutrino masses and mixing   in this framework.  In Sec. \ref{sec:results} numerical correlational study between observables of neutrino sector and model input parameters is established. We also present a brief  discussion of the non-unitarity effect and lepton flavor violation. 
Leptogenesis in the context of the present model is discussed in Sec. \ref{sec:lepto}
and a brief discussion on collider signature is presented in Sec. \ref{collider}. Finally in Sec. \ref{sec:con},  we conclude our results.

\section{MODEL FRAMEWORK}
\label{sec:linear}
This model represents the simplistic scenario of linear seesaw, where the particle content and group charges are provided in Table \ref{tab:fields-linear}. We prefer to extend with discrete $A_4$ modular symmetry to explore the neutrino phenomenology and a global $U(1)_X$ symmetry 
is imposed to forbid certain unwanted terms in the superpotential. The particle spectrum is enriched with six extra singlet heavy fermion superfields ($N_{Ri}$ and $S_{Li}$) and one weighton field ($\rho$). The extra supermultiplets of the model transform as triplet under the $A_4$ modular group. The $A_4$ and $U(1)_X$ symmetries are considered to be broken at a scale much higher than the electroweak symmetry breaking \cite{Dawson:2017ksx}. The extra superfields acquire masses by assigning non-zero vacuum expectation value  to the singlet weighton. The modular weight is assigned to all the particles and denoted as $k_I$. Further, it is evident that the breaking of $U(1)_X$ symmetry takes places by singlet $\rho$ acquiring VEV. Therefore, a massless Goldstone boson comes into picture which does not have dangerous interaction among the SM particles but interact only with Higgs and contributes to the dark radiation \cite{Lindner:2011it,Garcia-Cely:2013nin}. The importance of $A_4$ modular symmetry is the requirement of less number of flavon or weighton fields unlike the usual $A_4$ group, since the Yukawa couplings have the non-trivial group transformation. Assignment of group charge and modular weight to the Yukawa coupling is provided in Table \ref{tab:coupling}.

\begin{center} 
\begin{table}
\centering
\begin{tabular}{|c||c|c|c|c|c|c||c|c|c|c|}\hline\hline  

Fields& ~$e^c_R$~& ~$\mu^c_R$~  & ~$\tau^c_R$~& ~${L}_L$~& ~$N_R$~& ~$S_L^c$~& ~$H_{u,d}$~&~$\rho$ \\ \hline 
$SU(2)_L$ & $1$  & $1$  & $1$  & $2$  & $1$  & $1$  & $2$   & $1$     \\\hline 
$U(1)_Y$   & $1$ & $1$ & $1$ & $-\frac12$  & $0$ & $0$  & $\frac12, -\frac12$  & $0$   \\\hline
$U(1)_X$   & $1$ & $1$ & $1$   &$-1$  & $1$  & $-2$  & $0$  &$1$  \\\hline
$A_4$ & $1$ & $1'$ & $1''$ & $1, 1^{\prime \prime}, 1^{\prime }$ & $3$ & $3$ & $1$ & $1$  \\ \hline

$k_I$ & $1$ & $1$ & $1$ & $-1$ & $-1$ & $-1$ & $0$  & $0$\\ 
\hline
\end{tabular}
\caption{Particle content of the model and their charges under $SU(2)_L\times U(1)_Y\times A_4$ where $k_I$ is the modular weight.}
\label{tab:fields-linear}
\end{table}
\end{center}
\begin{center} 
\begin{table}
\centering
\begin{tabular}{|c||c|c|c|c|c|}\hline
{Yukawa coupling}  & ~{ $A_4$}~& ~$k_I$~     \\\hline 
{ $ \rm Y$} & ${\bf 3}$ & ${\bf 2}$      \\\hline
\end{tabular}
\caption{Modular weight of the Yukawa coupling $\bm{Y}$ and its transformation under $A_4$ symmetry.}
\label{tab:coupling}
\end{table}
\end{center}
\subsection{ Dirac mass term for charged leptons ($M_{\ell}$)}

In order to have a simplified structure for charged leptons mass matrix, we consider the three generations of left-handed doublets ($L_{e_L}, L_{\mu_L}, L_{\tau_L} $) transform as $\bm{1}, \bm{1}^{\prime \prime}, \bm{1}^{\prime}$ respectively under the $A_4$ symmetry. They are assigned the $U(1)_X$ charge of $-1$ for each generation. The right-handed charged leptons follow a transformation of $\bm{1}, \bm{1}^{ \prime}, \bm{1}^{\prime \prime}$ under $A_4$ and singlets in $U(1)_X$ symmetries respectively. All of them are assigned with a modular weight of 1. The VEVs of Higgs superfields i.e. $\langle H_u\rangle = v_u/\sqrt2, \langle H_d\rangle =v_d/\sqrt2$ are related to SM Higgs VEV as $v_H = \sqrt{v^2_u + v^2_d}$ and the ratio of their VEVs is expressed as $\tan\beta= ({v_u}/{v_d})=5$ \cite{Antusch:2013jca,Kashav:2021zir}. The relevant superpotential term for charged leptons is given by
\begin{align}
 \mathcal{W}_{M_\ell}  
                   &= y_{\ell_{}}^{ee}  {L}_{e_L} H_d ~e_R^c +  y_{\ell_{}}^{\mu \mu}  {L}_{\mu_L} H_d~ \mu_R^c +  y_{\ell_{}}^{\tau \tau}  {L}_{\tau_L} H_d~ \tau_R^c\;.
                    \label{Eq:yuk-Mell} 
\end{align}
The charged lepton mass matrix is found to be diagonal and the couplings can be adjusted to achieve the observed charged lepton masses. The mass matrix takes the form
\begin{align}
M_\ell = \begin{pmatrix}  y_{\ell_{}}^{ee} v_d/\sqrt{2}  &  0 &  0 \\
                                       0  &  y_{\ell_{}}^{\mu \mu} v_d/\sqrt{2}  &  0 \\
                                       0  &  0  &  y_{\ell_{}}^{\tau \tau} v_d/\sqrt{2}        \end{pmatrix}  =
                     \begin{pmatrix}  m_e  &  0 &  0 \\
                                       0  &  m_\mu  &  0 \\
                                       0  &  0  &  m_\tau      \end{pmatrix}.                 
\label{Eq:Mell} 
\end{align}
Here, $m_e$, $m_\mu$ and $m_\tau$ are the observed charged lepton masses.

\subsection{Dirac and pseudo-Dirac mass terms for the light neutrinos}
Along with the transformation of lepton doublets mentioned previously, the right-handed fermion superfields transform as triplets under $A_4$ modular group with  $U(1)_X$ charge of 1 and modular weight $-1$. Since, with these charge assignments we can not write the standard interaction  term, we introduce the Yukawa couplings to transform non-trivially under the $A_4$ modular group (triplets) and assign with modular weight of 2, as represented in Table\,\ref{tab:coupling}. We use the modular forms of the coupling as {$\bm{Y}(\tau) = \left(y_{1}(\tau),y_{2}(\tau),y_{3}(\tau)\right)$}, which can be written in terms of Dedekind eta-function  $\eta(\tau)$ and its derivative \cite{Feruglio:2017spp}, expressed in Eq.~\eqref{eq:Y-A4} (Appendix). Therefore, the invariant Dirac superpotential involving the active and right-handed fermion superfields can be written as
\begin{align}
 \mathcal{W}_{D}  
                   &= \alpha_D   {L}_{e_L} H_u~ (\bm{Y} N_R)_{1}   + \beta_D   {L}_{\mu_L} H_u~ (\bm{Y} N_R)_{1^{\prime}}
                   + \gamma_D   {L}_{\tau_L} H_u~ (\bm{Y} N_R)_{1^{\prime \prime}}.                       
                   \label{Eq:yuk-MD} 
\end{align}
Here, the subscript for the operator $\bm{Y} N_R$ indicates $A_4$ representation constructed by the product and $\{\alpha_D, \beta_D, \gamma_D\}$ are free parameters. The resulting Dirac neutrino mass matrix is found to be
\begin{align}
M_D&=\frac{v_u}{\sqrt2}
\left[\begin{array}{ccc}
\alpha_D & 0 & 0 \\ 
0 & \beta_D & 0 \\ 
0 & 0 & \gamma_D \\ 
\end{array}\right]
\left[\begin{array}{ccc}
y_1 &y_3 &y_2 \\ 
y_2 &y_1 &y_3 \\ 
y_3 &y_2 &y_1 \\ 
\end{array}\right]_{LR}.                   
\label{Eq:Mell} 
\end{align}
As we also have the extra sterile fermion superfields $S_{Li}$, which transform analogous to $N_{Ri}$ under $A_4$ modular symmetry, the pseudo-Dirac term for the light neutrinos is allowed, and the corresponding  superpotential is given as 
\begin{align}
 \mathcal{W}_{LS}  
                   &= \Big[\alpha'_D   {L}_{e_L} H_u~ (\bm{Y} S^c_L)_{1}   + \beta'_D   {L}_{\mu_L} H_u~ (\bm{Y} S^c_L)_{1^{\prime}}
                   + \gamma'_D   {L}_{\tau_L} H_u~ (\bm{Y} S^c_L)_{1^{\prime \prime}}\Big] \frac{\rho^3}{\Lambda^3} \;,
                   \label{Eq:yuk-LS} 
\end{align}
where, the subscript for the operator $(\bm{Y} S^c_L)$ indicates $A_4$ representation constructed by the product and $\{\alpha'_D, \beta'_D, \gamma'_D\}$ are free parameters. The flavor structure for the pseudo-Dirac neutrino mass matrix takes the form,
\begin{align}
M_{LS}&=\frac{v_u}{\sqrt2}\left(\frac{v_\rho}{\sqrt{2}\Lambda}\right)^3
\left[\begin{array}{ccc}
\alpha^\prime_D & 0 & 0 \\ 
0 & \beta^\prime_D & 0 \\ 
0 & 0 & \gamma^\prime_D \\ 
\end{array}\right]
\left[\begin{array}{ccc}
y_1 &y_3 &y_2 \\ 
y_2 &y_1 &y_3 \\ 
y_3 &y_2 &y_1 \\ 
\end{array}\right]_{LR}.                   
\label{Eq:Mell} 
\end{align}\\
\subsection{ Mixing between the heavy superfields $N_R$ and $S_L^c$}
Following the transformation of the heavy fermion superfields under the imposed symmetries, it can be noted that the usual Majorana mass terms are not allowed. But one can have the interactions leading to the mixing between these additional superfields as follows
\begin{eqnarray}
 \mathcal{W}_{M_{RS}}  
                   &=& [\alpha_{NS} \bm{Y} ({S^c_L} N_R)_{\rm sym} + \beta_{NS} \bm{Y} ({S^c_L} N_R)_{\rm Anti-sym} ]\rho  
                    \nn \\
                   &=&\alpha_{NS}[ y_1(2  S^c_{L_1} N_{R_1} - S^c_{L_2} N_{R_3} - S^c_{L_3} N_{R_2})+y_2(2   S^c_{L_2} N_{R_2} - S^c_{L_1} N_{R_3} - S^c_{L_3} N_{R_1}) \nn \\
&+& y_3(2   S^c_{L_3} N_{R_3} -  S^c_{L_1} N_{R_2} - S^c_{L_2} N_{R_1})] \rho \nn\\
&+&\beta_{NS}[ y_1(  S^c_{L_2} N_{R_3} - S^c_{L_3} N_{R_2})+y_2(  S^c_{L_3} N_{R_1} -  S^c_{L_1} N_{R_3})+
y_3(   S^c_{L_1} N_{R_2} -  S^c_{L_2} N_{R_1})] \rho\;,  \nn\\
     \label{Eq:yuk-M} 
\end{eqnarray}
where the first and second terms in the first line correspond to symmetric and anti-symmetric product for $S^c_L N_R$ making triplet representation of $A_4$ with $\alpha_{NS}$, $\beta_{NS}$ being the free parameters.
Using $\langle \rho \rangle = v_\rho/\sqrt{2}$,  the resulting mass matrix is found to be,

\begin{align}
M_{RS}&=\frac{v_\rho}{\sqrt2}
 \left(
 \frac{\alpha_{NS}}{3}\left[\begin{array}{ccc}
2y_1 & -y_3 & -y_2 \\ 
-y_3 & 2y_2 & -y_1 \\ 
-y_2 & -y_1 & 2y_3 \\ 
\end{array}\right]
+
\beta_{NS}
\left[\begin{array}{ccc}
0 &y_3 & -y_2 \\ 
-y_3 & 0 & y_1 \\ 
y_2 & -y_1 &0 \\ 
\end{array}\right]
\right). \label{yuk:MRS}
\end{align}
 It should be noted that $\alpha_{NS}/3 \neq \beta_{NS}$, otherwise the matrix $M_{RS}$  becomes singular, which eventually spoils the intent of linear seesaw. The masses for the heavy fermions can be found in the basis $( N_R, S_L^c)^T$, which can be written as
\begin{eqnarray}
M_{Hf}= \begin{pmatrix}
0 & M_{RS}\\
M^T_{RS} & 0
\end{pmatrix}.\label{mrs matrix}
\end{eqnarray}
Therefore, one can have six doubly degenerate mass eigenstates for the heavy superfields upon diagonalization.

\subsection{Linear Seesaw mechanism for light neutrino Masses}
Within the present model invoked with $A_4$ modular symmetry, the complete $9 \times 9$ mass matrix  in the flavor basis of $\left(\nu_L, N_R, S^c_L \right)^T$ is given by
\begin{eqnarray}
\mathbb{M} = \left(\begin{array}{c|ccc}   & \nu_L & N_R  & S^c_L   \\ \hline
\nu_L  & 0       & M_D       & M_{LS} \\
N_R    & M^T_D         & 0       & M_{RS} \\
S^c_L & M_{LS}^T     & M_{RS}^T    & 0
\end{array}
\right).
\label{eq:numatrix-complete}
\end{eqnarray}
The linear seesaw mass formula for light neutrinos is given with the assumption $ M_{RS} \gg M_D, M_{LS}$ as,
\begin{eqnarray}
m_\nu 
&=& M_D M_{RS}^{-1} M_{LS}^{T}+{\rm transpose}.\label{mass}
\end{eqnarray}
Apart from the small neutrino masses,  other relevant parameters in the neutrino sector are Jarlskog invariant and the effective neutrino mass which play a key role in  neutrinoless double beta decay and  can be computed from the mixing angles and phases of PMNS matrix elements as following:
\begin{eqnarray}
&& J_{CP} = \text{Im} [U_{e1} U_{\mu 2} U_{e 2}^* U_{\mu 1}^*] = s_{23} c_{23} s_{12} c_{12} s_{13} c^2_{13} \sin \delta_{CP},\\
&& \left | m_{ee}\right |=|m_{1} \cos^2\theta_{12} \cos^2\theta_{13}+ m_{2} \sin^2\theta_{12} \cos^2\theta_{13}e^{i\alpha_{21}}+  m_{3} \sin^2\theta_{13}e^{i(\alpha_{31}-2\delta_{CP})}|\;.
\end{eqnarray}
Many dedicated experiments are looking for neutrinoless double beta signals,  for details please refer to \cite{Giuliani:2019uno}. The sensitivity limits  on $\left | m_{ee}\right |$ by the current experiments  such as GERDA  is $(102-213)$ meV \cite{Agostini:2019hzm} and CUORE is $(90-420)$ meV  \cite{Alduino:2017ehq}.
The future generation experiments, like LEGEND-200 can probe 35-73 meV \cite{Giuliani:2019uno} and  KamLAND-Zen  $(61-165)$ meV \cite{KamLAND-Zen:2016pfg}.  
\section{NUMERICAL ANALYSIS}
\label{sec:results}
\begin{figure}[h!]
\begin{center}
\includegraphics[height=50mm,width=75mm]{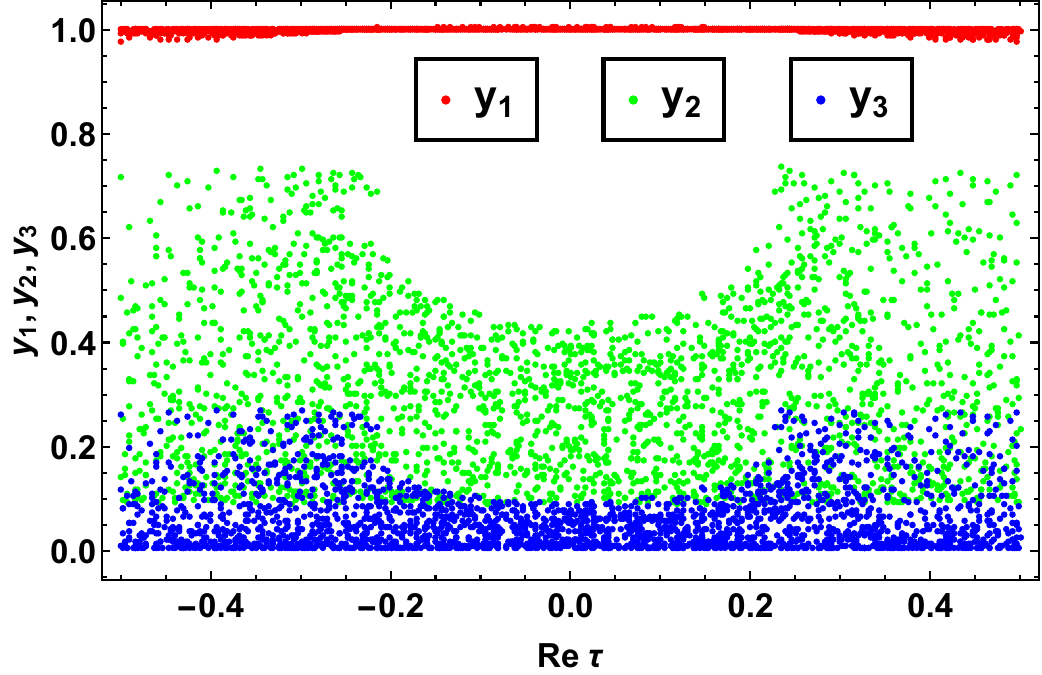}
\hspace*{0.2 true cm}
\includegraphics[height=50mm,width=75mm]{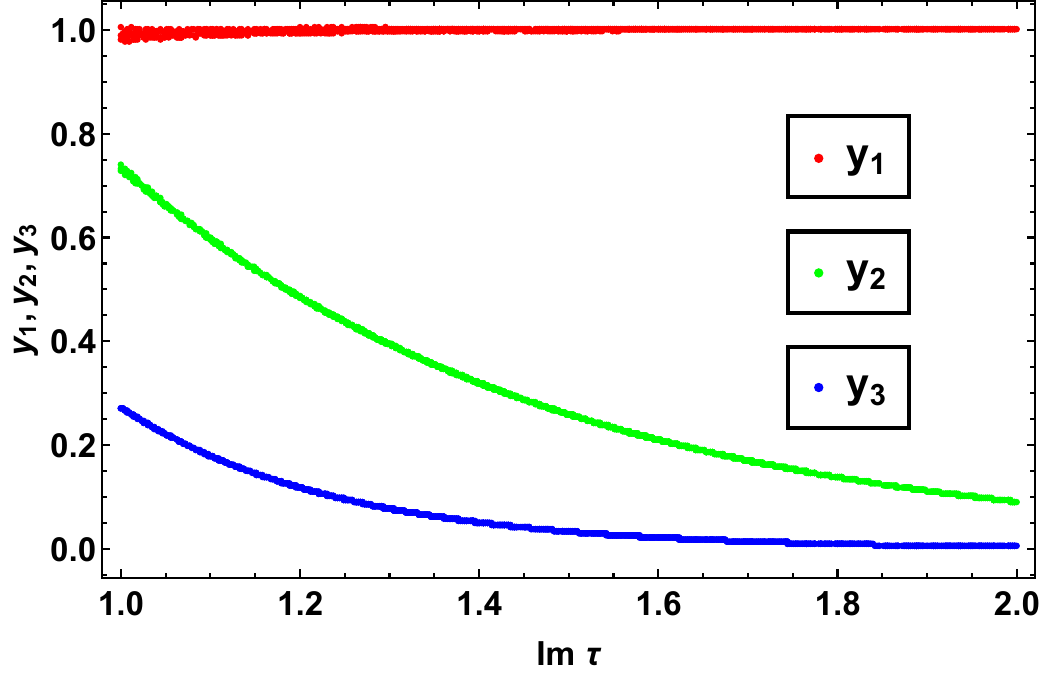}
\includegraphics[height=50mm,width=75mm]{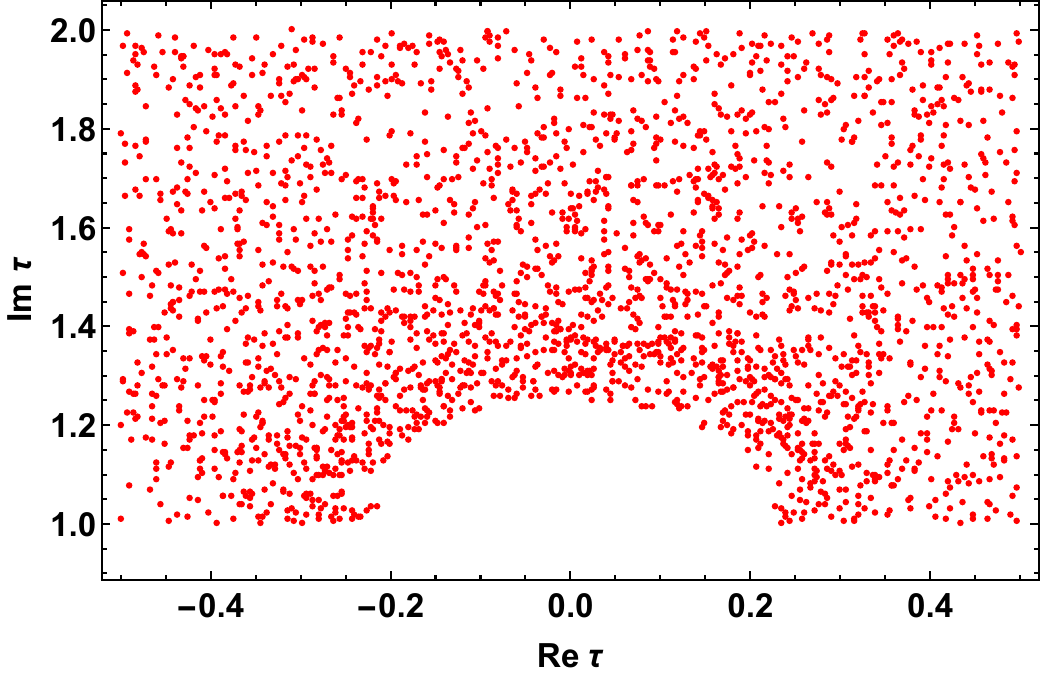}
\caption{Top left and top right panel signify the correlation of the modular Yukawa couplings ($y_1,y_2,y_3$) with the real and imaginary parts of modulus $\tau$ respectively. The bottom panel represents the allowed region of the Re($\tau$) and Im($\tau$) abiding all the constraints and within the range of its fundamental domain. }
\label{yuk_reim_tau}
\end{center}
\end{figure}

For numerical analysis we consider the global fit neutrino oscillation data at 3$\sigma$ interval from~\cite{Esteban:2018azc} as follows:
\begin{align}
&{\rm NO}: \Delta m^2_{\rm atm}=[2.431, 2.622]\times 10^{-3}\ {\rm eV}^2,\
\Delta m^2_{\rm sol}=[6.79, 8.01]\times 10^{-5}\ {\rm eV}^2,\nn\\
&\sin^2\theta_{13}=[0.02044, 0.02437],\ 
\sin^2\theta_{23}=[0.428, 0.624],\ 
\sin^2\theta_{12}=[0.275, 0.350].\label{eq:mix}
\end{align} 
Here, we numerically diagonalize the neutrino mass matrix eqn.(\ref{mass}) through the relation $U^\dagger {\cal M}U= {\rm diag}(m_1^2, m_2^2, m_3^2)$, where  ${\cal M}=m_\nu m_\nu^\dagger$ and $U$ is an unitary matrix, from which the neutrino mixing angles can be extracted  using the standard relations:
\begin{eqnarray}
\sin^2 \theta_{13}= |U_{13}|^2,~~~~\sin^2 \theta_{12}= \frac{|U_{12}|^2}{1-|U_{13}|^2}\;,~~~~~\sin^2 \theta_{23}= \frac{|U_{23}|^2}{1-|U_{13}|^2}\;.
\end{eqnarray}
To fit to the current neutrino oscillation data, we chose the following ranges for  the model parameters:
\begin{align}
&{\rm Re}[\tau] \in [-0.5,0.5],~~{\rm Im}[\tau]\in [1,2],~~ \{ \alpha_{D},\beta_{D},\gamma_D \} \in 10^{-5}~[0.1,1],~~\{ \alpha^\prime_{D},\beta^\prime_{D},\gamma^\prime_D \} \in 10^{-2}~[0.1,1], \nn \\
& \quad \alpha_{NS}  \in [0,0.5],\quad \beta_{NS} \in[0,0.0001],\quad v_\rho \in \nn [10,100] \ {\rm TeV},  \quad \Lambda \in  [100,1000] \ {\rm TeV}.
\end{align}
The input parameters are randomly scanned over the above mentioned ranges and the allowed regions for those are initially filtered by the observed $3\sigma$ limit of solar and atmospheric mass squared differences and mixing angles which are further constrained by the observed sum of active neutrino masses $\sum m_i < 0.12$ eV \cite{Aghanim:2018eyx}.
The typical range of modulus $\tau$ is found to be $-0.5\ \lesssim\ $Re$[\tau]\lesssim$\ 0.5 and  1\ $\lesssim\ $Im$[\tau]\lesssim$\ 2 for normally ordered neutrino masses. Thus, the modular Yukawa couplings as function of $\tau$ (Eq. \eqref{eq:Y-A4} in Appendix) are found to vary in the region 0.99\ $\lesssim\ $$y_1$$(\tau)\lesssim$\ 1,  0.1\ $\lesssim\ $$y_2$$(\tau)\lesssim$\ 0.8 and 0.01\ $\lesssim\ $$y_3$$(\tau)\lesssim$\ 0.3. The variation of those Yukawa couplings with the real and imaginary parts of $\tau$ are represented in the top left and top right panels of Fig. \ref{yuk_reim_tau} respectively, whereas, bottom panel shows the allowed region of Re($\tau$) and Im($\tau$) which abides all the constraints used to deduce the neutrino oscillation parameters. 

\begin{figure}[h!]
\begin{center}
\includegraphics[height=50mm,width=75mm]{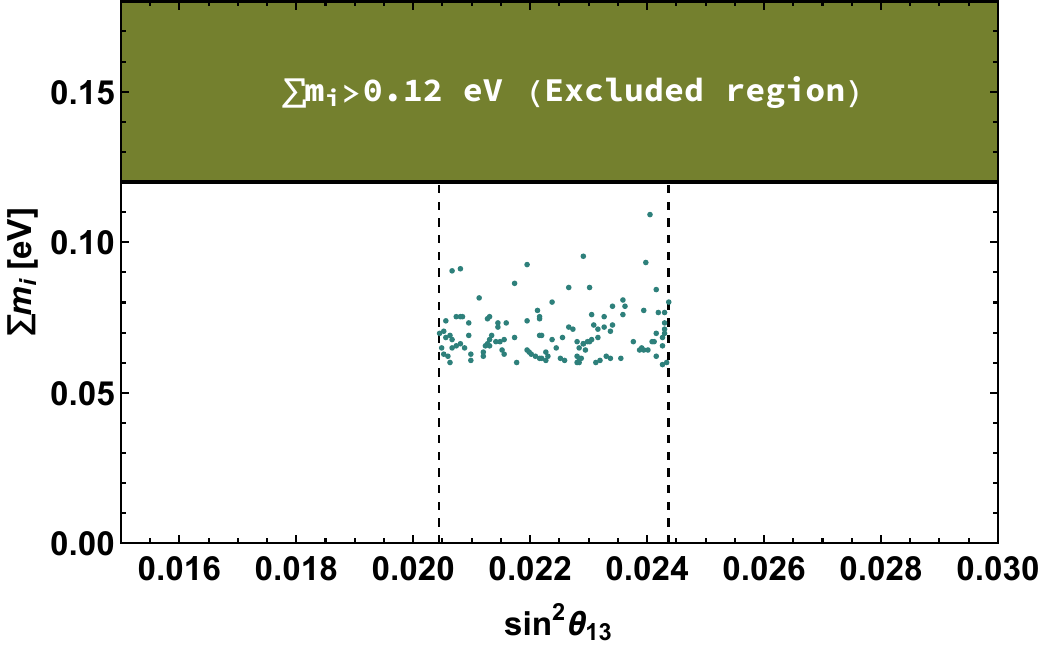}
\includegraphics[height=50mm,width=75mm]{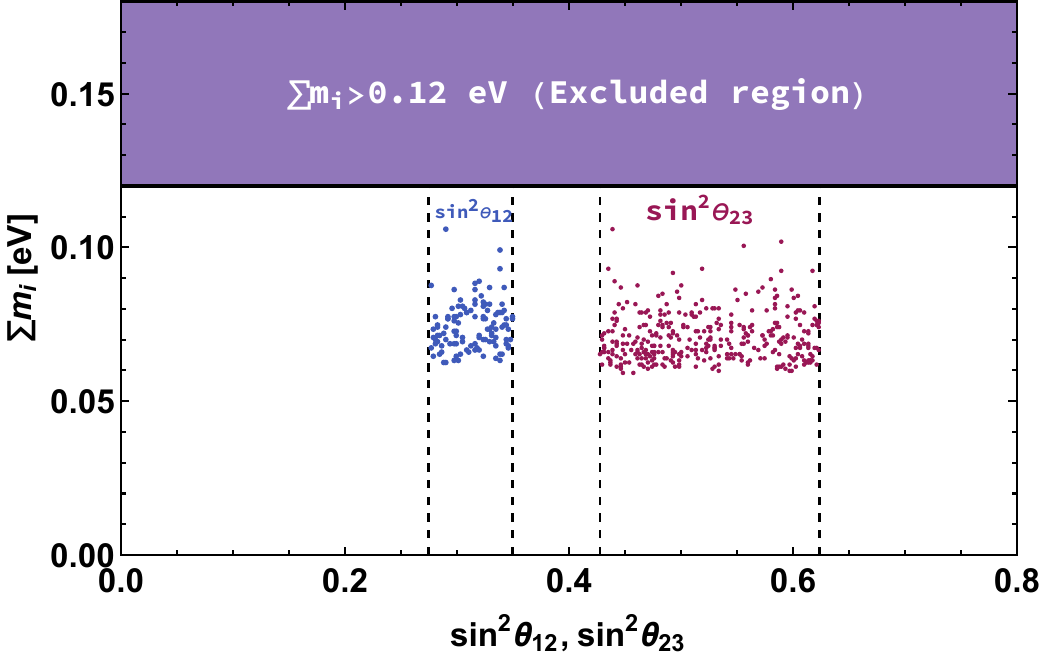}\\
\caption{ Left (Right) panel represents the correlation between $\sin^2 \theta_{13}$ ($\sin^2\theta_{12}$ and $\sin^2\theta_{23}$) with the sum of active neutrino masses.}
\label{mix_angles}
\end{center}
\end{figure}

\begin{figure}[h!]
\begin{center}
\includegraphics[height=50mm,width=75mm]{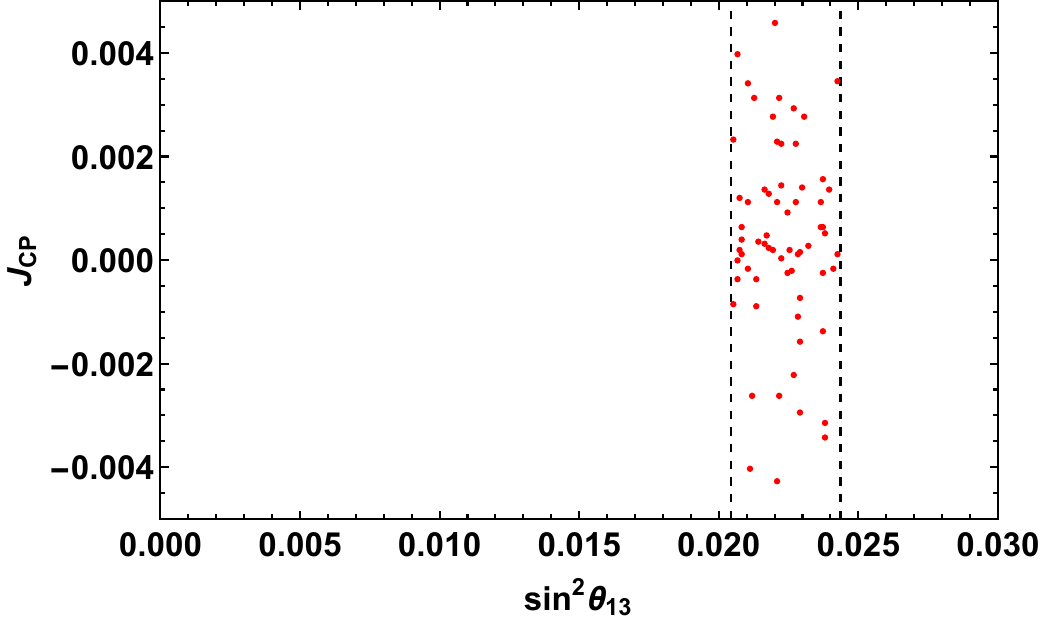}
\hspace*{0.2 true cm}
\includegraphics[height=50mm,width=75mm]{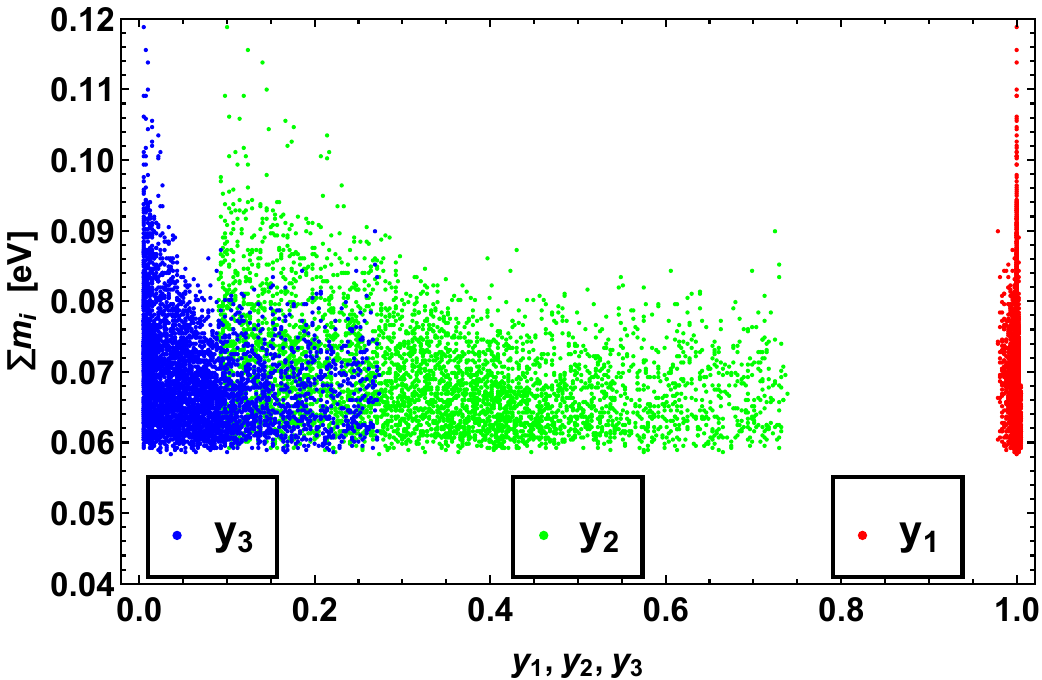}
\caption{Left panel displays the correlation of Jarsklog invariant with the reactor mixing angle and right panel reflects the variation of modular Yukawa couplings with the sum of active neutrino masses.}
\label{y_jcp}
\end{center}
\end{figure}

\begin{figure}[h!]
\begin{center}
\includegraphics[height=50mm,width=75mm]{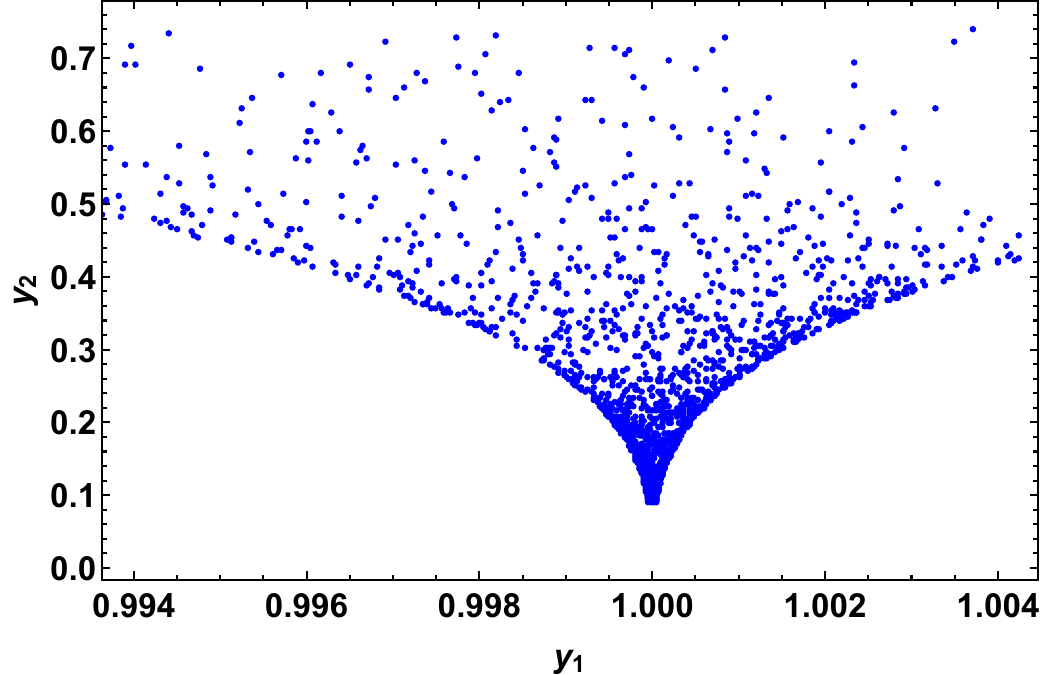}
\hspace*{0.2 true cm}
\includegraphics[height=50mm,width=75mm]{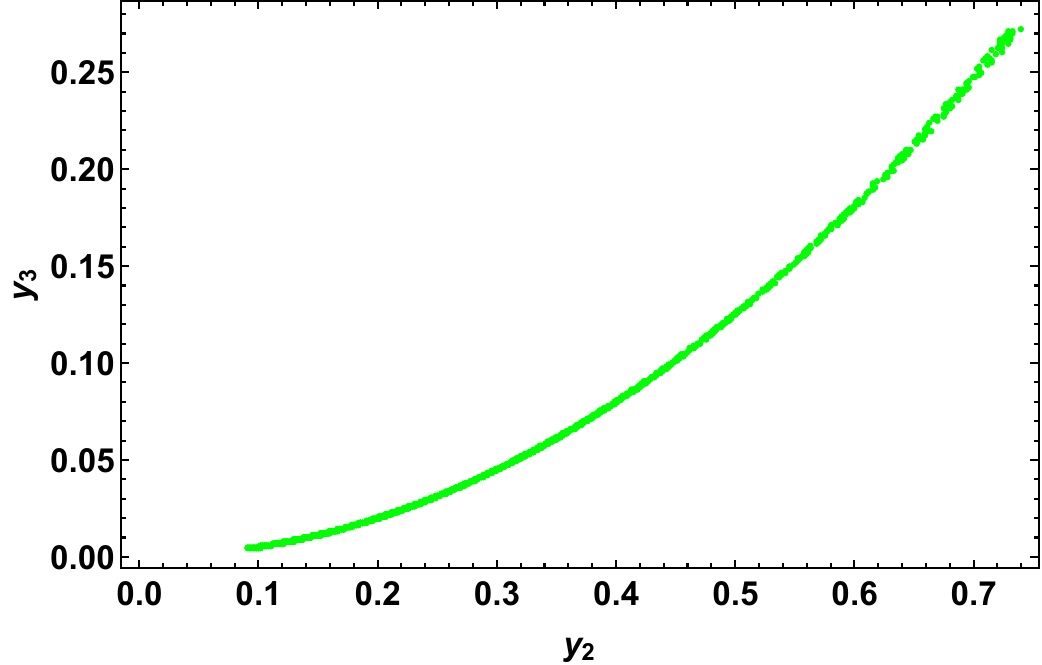}
\caption{Left (Right) panel displays the correlation between $y_1$ and $y_2$ ($y_2$ and $y_3$).}
\label{y1_y2_y3}
\end{center}
\end{figure}

\begin{figure}[h!]
\begin{center}
\includegraphics[height=50mm,width=75mm]{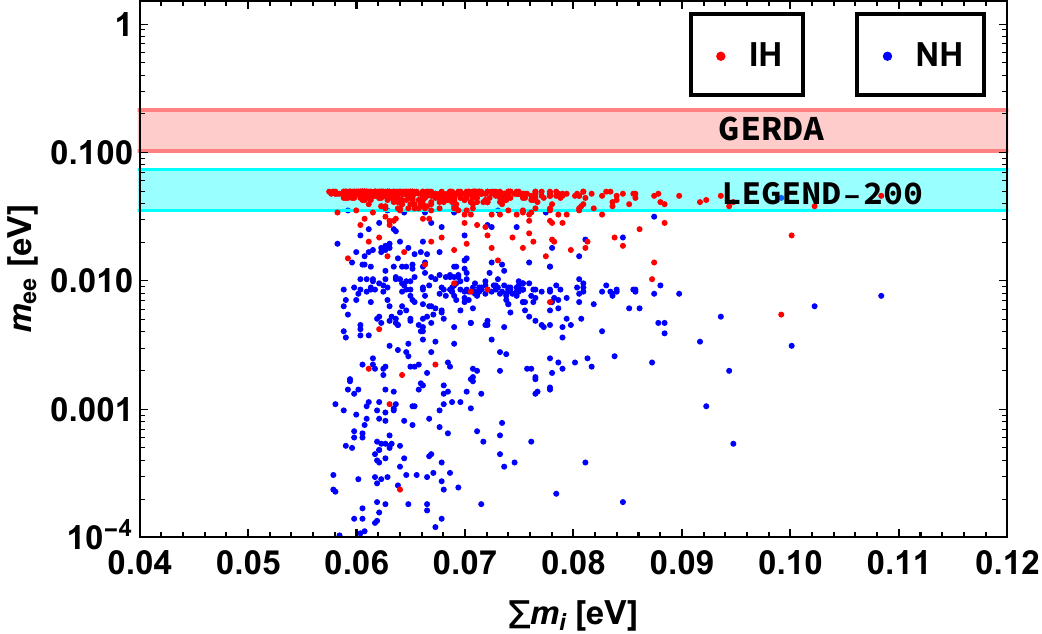}
\hspace*{0.2 true cm}
\includegraphics[height=50mm,width=72mm]{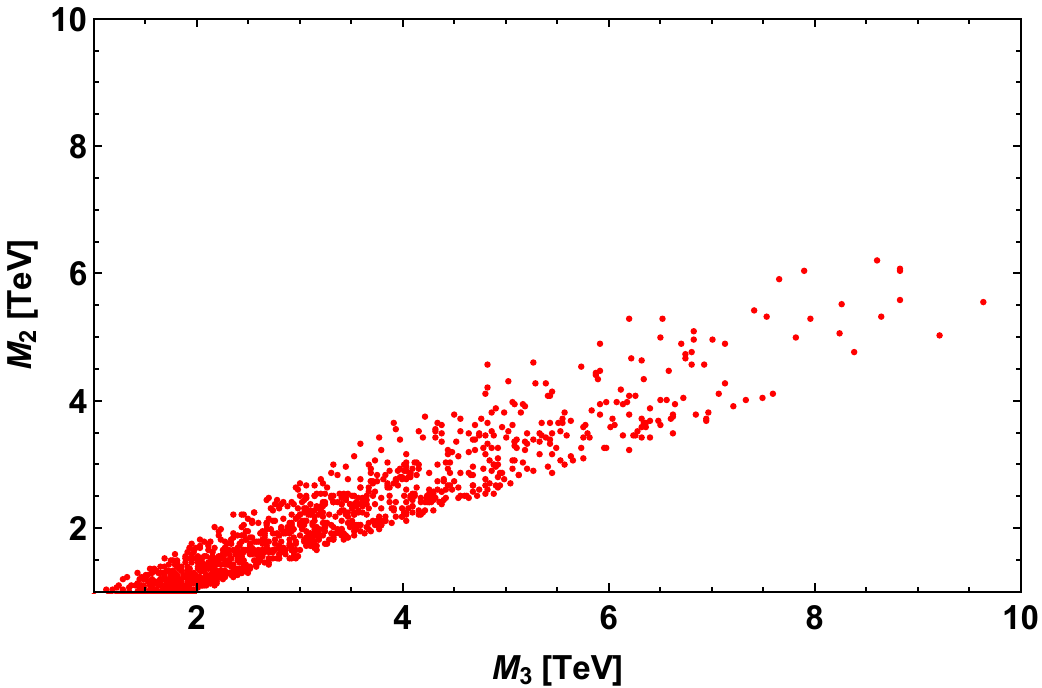}
\caption{Left panel shows   the correlation of effective neutrino mass of neutrinoless double beta decay with the sum of active neutrino masses, where the blue and red points correspond to normal and inverted hierarchies. The horizontal pink band corresponds to the $3\sigma$ sensitivity limit of currently running GERDA experiment and the cyan band represents the $3\sigma$ limit of the future LEGEND-200 experiment. Right panel depicts correlation between the heavy fermion masses $M_2$ and $M_3$.}
\label{M23_mee}
\end{center}
\end{figure}

Variation of the mixing angles with the sum of active neutrino masses, consistent  with the allowed $3\sigma$ range are obtained, as shown in Fig. \ref{mix_angles}. In the left panel of Fig. \ref{y_jcp}, we show the correlation of Jarsklog CP invariant with the reactor mixing angle allowed by the neutrino oscillation data, which is found to be of the order of ${\cal O}(10^{-3})$. The right panel of Fig. \ref{y_jcp}, signifies the full parameter space for Yukawa couplings as per the observed sum of active neutrino masses. In Fig. \ref{y1_y2_y3}, we have displayed a correlation of the Yukawa couplings $y_1$ with $y_2$ and $y_2$ with $y_3$ in the left and right panels respectively. The effective  neutrinoless double beta decay mass parameter $|m_{ee}|$ for both normal and inverted orderings is found to have a maximum value of $55$ meV from the variation of observed sum of active neutrino masses, which is presented in the left panel of Fig. \ref{M23_mee}.  The results for  normal and inverted hierarchies are shown by the blue and red points. The horizontal pink and cyan bands represent the $3\sigma$ sensitivity limits of current GERDA  and the future LEGEND-200 experiments respectively. It should be noted from the figure that the model predictions for $|m_{ee}|$ are within the reach of the future generation experiments and the inverted hierarchical region is more favoured. The right panel represents the correlation between heavy fermion masses $M_2$ and $M_3$. 

\section*{Comment on non-unitarity}
\label{sec:non-unitarity}
Here, we briefly comment on non-unitarity of neutrino mixing matrix $U'_{\rm PMNS}$ in the presence of heavy fermionic superfields.
The standard parametrization for the deviation from unitarity  can be expressed as   \cite{Forero:2011pc}
\begin{align}
U'_{\rm PMNS}\equiv \left(1-\frac12 FF^\dag\right) U_{\rm PMNS}.
\end{align}
Here,  $U_{\rm PMNS}$ is the PMNS mixing matrix which diagonalises the mass matrix of the three light neutrinos and $F$ is the mixing of active neutrinos with the heavy fermions and approximated as $F\equiv  (M^{T}_{RS})^{-1} M_D \approx \frac{\alpha_D v}{\alpha_{NS} v_\rho}$, which is a hermitian matrix.
The global constraints on the non-unitarity parameters  \cite{Antusch:2014woa,Blennow:2016jkn,Fernandez-Martinez:2016lgt}, are found via several experimental results such as the $W$ boson mass $M_W$, the  Weinberg angle $\theta_W$, several ratios of $Z$ boson fermionic decays as well as its  invisible decay, electroweak universality,  CKM unitarity bounds, and lepton flavor violations. In our model framework, we consider the following approximated normalized order for the Dirac, pseudo-Dirac and heavy fermion masses to correctly generate the observed mass-squared differences as well as the  sum of active neutrino masses of desired order
\begin{eqnarray}
\left(\frac{m_\nu}{0.1 ~{\rm eV}} \right) \approx \left(\frac{M_D}{10^{-3}~~ {\rm GeV}}\right) \left(\frac{M_{RS}}{10^3~~ {\rm GeV}}\right)^{-1} \left( \frac{M_{LS}}{10^{-4}~~ {\rm GeV}}\right). 
\end{eqnarray}
Therefore, with the chosen order of masses, we obtain an approximated non-unitary mixing for the present model as

\begin{align}
|FF^\dag|\le  
\left[\begin{array}{ccc} 
6.06\times 10^{-13} & 4.25\times 10^{-14}  & 6.9\times 10^{-14}  \\
4.25\times 10^{-14}  & 2.77\times 10^{-13}  & 1.14\times 10^{-12}  \\
6.9\times 10^{-14}  & 1.14\times 10^{-12}  & 7.25\times 10^{-13} \\
 \end{array}\right].\label{non-uni}
\end{align}
Since, the mixing between the active and heavy fermions  in our model  is found to be very small,  it  leads to a negligible contribution to the non-unitarity.
\section*{Comment on lepton flavor violation}
\label{sec:non-LFV}
Here, we will briefly discuss about the prospect of  lepton flavor violation (LFV) effect, in particular $\ell_i \to \ell_j \gamma$ decays, in the context of present model. Lepton flavor violating decays are strictly forbidden in the SM and  are known to be  induced in models with extended lepton sectors. The current limit on  these branching ratios are: ${\rm Br}(\mu \to e \gamma) < 4.2 \times 10^{-13}$  from MEG Collaboration \cite{TheMEG:2016wtm},  ${\rm Br}(\tau \to e \gamma)<3.3\times 10^{-8}$ and ${\rm Br}(\tau \to \mu \gamma)<4.4 \times 10^{-8}$  from BABAR collaboration \cite{Aubert:2009ag}.

In this model,  the lepton flavor violating  decays ($\ell_i \to \ell_j \gamma$) can occur via exchange of heavy fermions at one loop level \cite{Bernabeu:1987gr,Deppisch:2004fa},  as there is mixing between the light and heavy fermions and the corresponding dominant one-loop contribution to the branching ratios for these decays is given as  \cite{Ilakovac:1994kj,Forero:2011pc}
\begin{equation}
\text{Br}(\ell_i \to \ell_j \gamma) = \frac{\alpha_W^3s_W^2}{256\pi^2}\frac{m_{\ell_i}^5}{M_W^4}\frac{1}{\Gamma_{\ell_i}}|G_{ij}^W|^2,
\end{equation}
where $G_{ij}^W$ is loop functions whose analytic form is
\begin{eqnarray}
G_{ij}^W &=& \sum_{k=1}^3 F_{ik} F_{jk}^\dagger G_{\gamma}^W \left (\frac{\ M_{N_k}^2}{M_W^2}\right )~~~~~\text{with}  \nonumber\\
G_{\gamma}^W (x)&=& \frac{1}{12(1-x)^4}(10-43x+78x^2-49x^3 +4x^4)\;.
\end{eqnarray}
Here, $M_{N_k}$ represents heavy  neutrino superfields and $F$ characterises the mixing of active neutrinos with the heavy fermions leading to non-unitarity effect. Since in the present model, the non-unitarity parameters are found to be extremely small (\ref{non-uni}),  the branching ratios of the LFV decays are highly suppressed. Thus, for  TeV scale heavy fermions $M_{N_k}$, the branching ratios  for different LFV decays are found to be
\bea
&&{\rm Br}(\mu \to e \gamma ) \leq 8.9 \times 10^{-33} \left (\frac{|(F F^\dagger)_{\mu e}|}{4.25 \times 10^{-14}} \right )^2,\nn\\
&&{\rm Br}(\tau \to e \gamma ) \leq 4.2 \times 10^{-33} \left (\frac{|(F F^\dagger)_{\tau e}|}{6.9 \times 10^{-14}} \right )^2,\nn\\
&&{\rm Br}(\tau \to \mu \gamma ) \leq1.2 \times 10^{-30} \left (\frac{|(F F^\dagger)_{\tau \mu }|}{1.14 \times 10^{-12}} \right )^2,
\eea
 which are beyond the reach of any of the  future experiments.


\section{Leptogenesis}
\label{sec:lepto}
\begin{figure}[h!]
\begin{center}
\includegraphics[height=48mm,width=53mm]{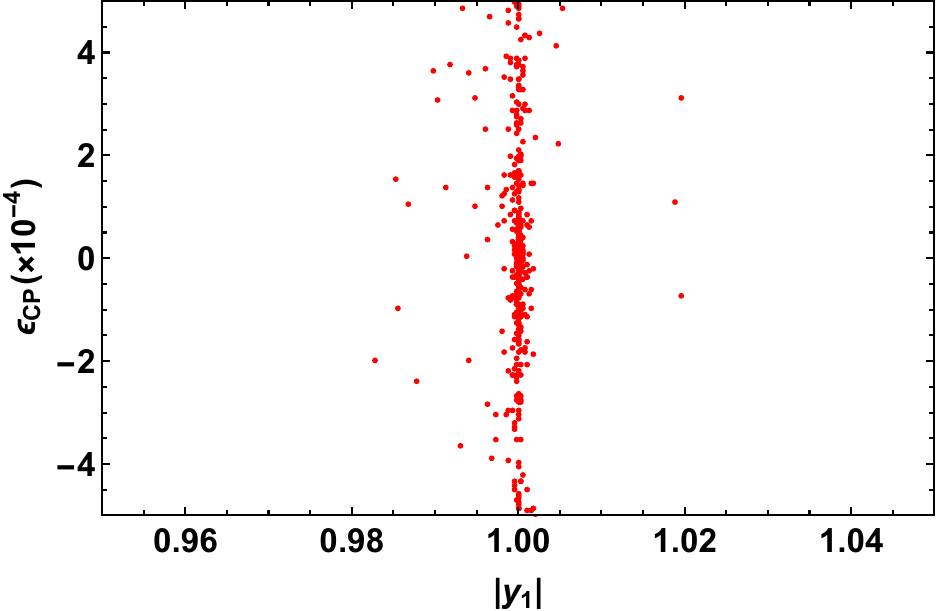}
\includegraphics[height=48mm,width=53mm]{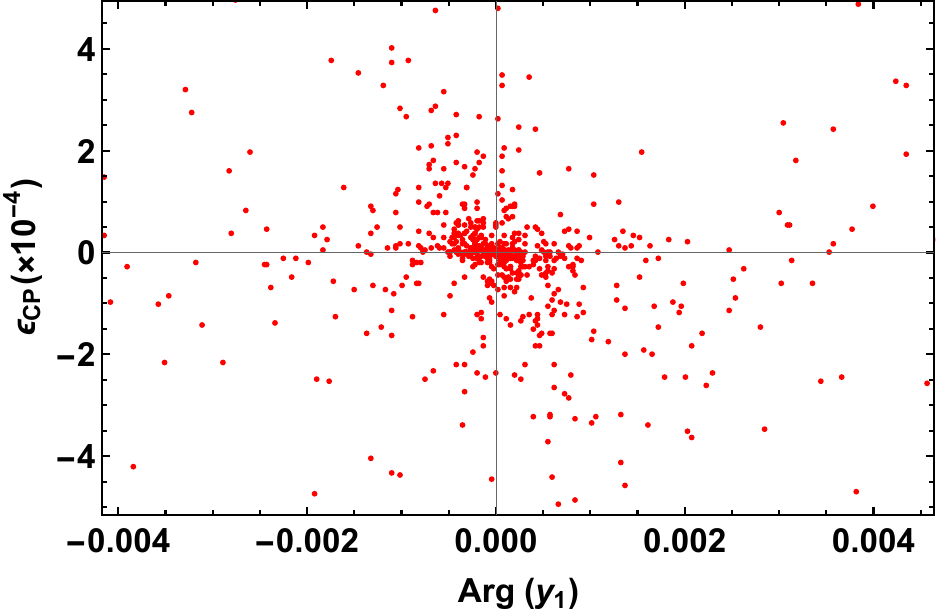}
\includegraphics[height=48mm,width=53mm]{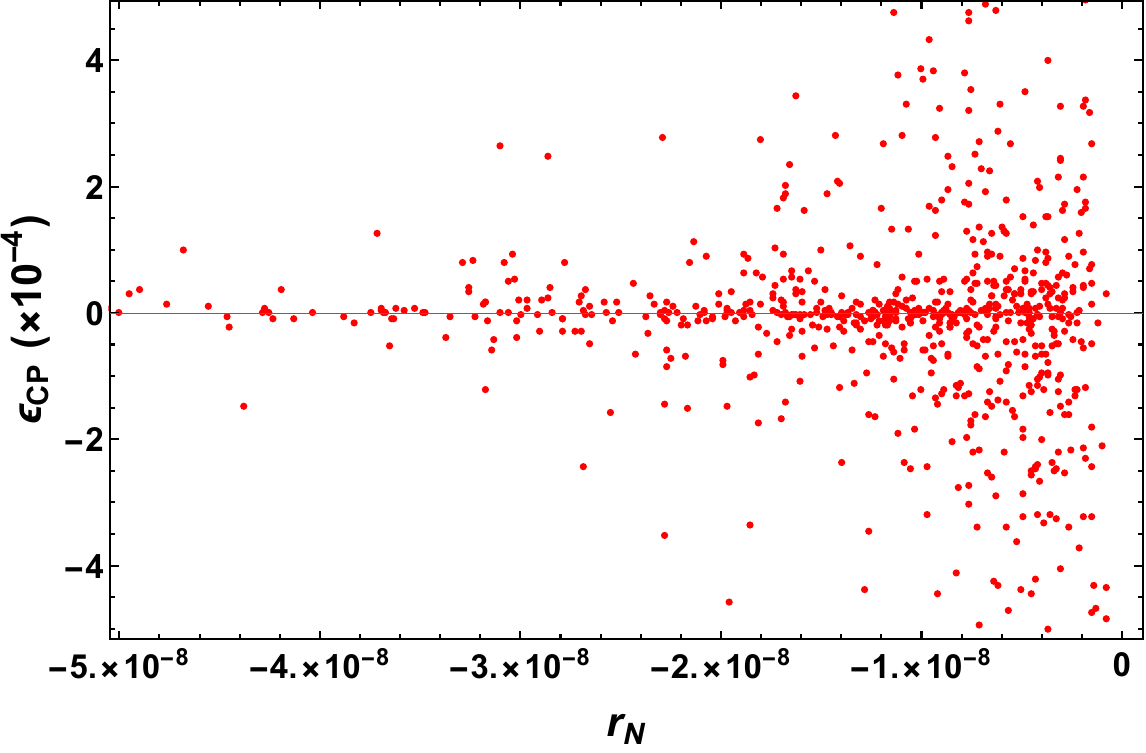}
\includegraphics[height=48mm,width=66mm]{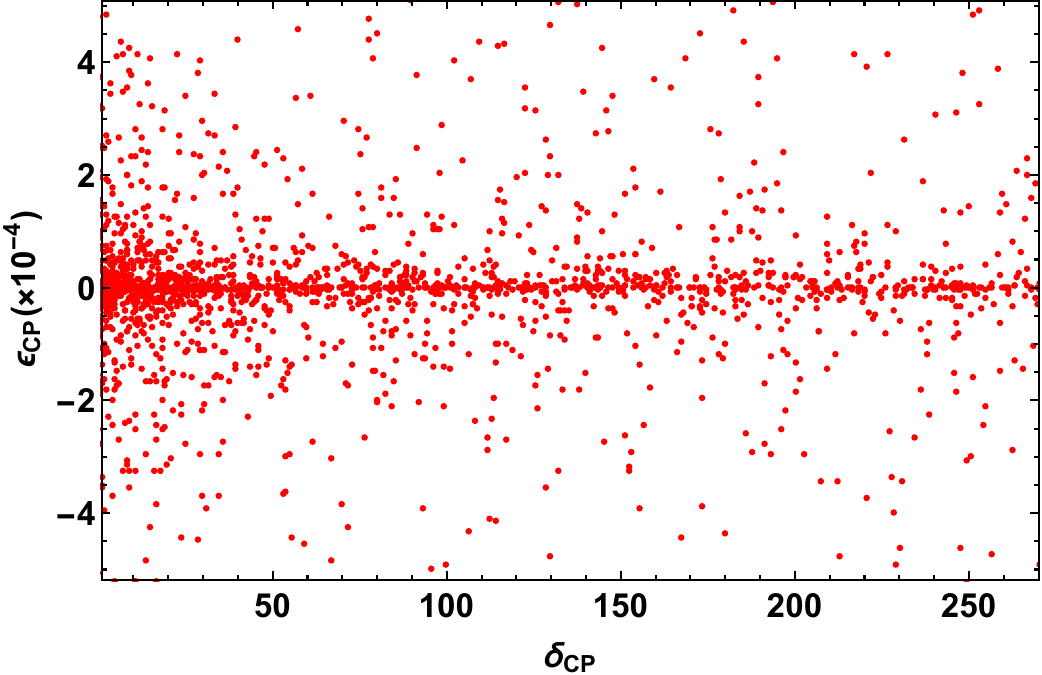}
\caption{Left and middle panels represent the variation of CP asymmetry with the magnitude and argument of Yukawa coupling respectively. Right panel shows its dependence with parameter $r_N$. Whereas, the bottom plot represents the correlation between CP asymmetry and  the CP violating phase $\delta_{CP}$. }
\label{CP_var}
\end{center}
\end{figure}

Leptogenesis has proven to be one of the most preferred way to generate the observed baryon asymmetry of the Universe. The standard scenario of resonant enhancement in CP asymmetry has brought down the scale as low as TeV \cite{Pilaftsis:1997jf,Bambhaniya:2016rbb,Pilaftsis:2003gt, Abada:2018oly}. The present model includes six heavy states  with doubly degenerate masses for each pair Eqn. \eqref{mrs matrix}. But one can introduce a higher dimensional mass term for the heavy neutrino superfield ($S_L^c$) as 
\begin{eqnarray}
L_M=-\alpha_R Y {S^c_L} S^c_L \frac{\rho^4}{\Lambda^3}.
\label{rhnmatrix}
\end{eqnarray} 
This leads to a small mass splitting between the heavy superfields, there by enhancing the CP asymmetry to generate required lepton asymmetry \cite{Pilaftsis:2005rv,Asaka:2018hyk}. 
Thus, one can construct the right-handed Majorana mass matrix as follows
\begin{equation}
M_R =\frac{\alpha_R v^4_\rho}{6\Lambda^3}\begin{pmatrix}
2y_1 & -y_3 & -y_2\\
-y_3 & 2y_2 & -y_1\\
-y_2 & -y_1 & 2y_3
\end{pmatrix}.
\end{equation}
The coupling $\alpha_R$ is chosen to be extremely small to retain the linear seesaw structure of the mass matrix Eqn. (\ref{eq:numatrix-complete}), i.e.,  $M_D, M_{LS} \gg M_R$ and such inclusion does not affect the previous results. However, this term introduces a small mass splitting and the $2 \times 2$ submatrix of Eqn. (\ref{eq:numatrix-complete}) in the $(N_R, S^c_L)$ basis, now can be written as
\begin{eqnarray}
M=\begin{pmatrix}
0 & M_{RS}\\
M_{RS}^T & M_R
\end{pmatrix}.
\end{eqnarray}
This matrix can have a block diagonal structure in the limit $\beta_{NS} \ll \alpha_{NS}$  by the unitary matrix $ \frac{1}{\sqrt 2}
\begin{pmatrix}
I & -I\\
I & I
\end{pmatrix} $ as
\begin{eqnarray}
M'=\begin{pmatrix}
  M_{RS}+\frac{M_R}{2} & -\frac{M_R}{2}\\
-\frac{M_R}{2} &  -M_{RS}+\frac{M_R}{2}
\end{pmatrix} \approx  \begin{pmatrix}
  M_{RS}+\frac{M_R}{2} & 0\\
0 &  -M_{RS}+\frac{M_R}{2}
\end{pmatrix}.
\end{eqnarray}\label{block-diag}
Therefore, the mass eigenstates ($N^\pm$) are related to $N_R$ and $S^c_L$ through
\begin{equation}
\begin{pmatrix}
S^c_{Li}\\N_{Ri}
\end{pmatrix}= \begin{pmatrix}
\cos{\theta} & -\sin{\theta}\\
\sin{\theta} & \cos{\theta}
\end{pmatrix} \begin{pmatrix}
N_i^+ \\ N_i^-
\end{pmatrix}.
\end{equation}   
Assuming a maximal mixing, we can have
\begin{eqnarray}
N_{Ri} = \frac{(N_i^+ + N_i^-)}{\sqrt{2}},~~ S^c_{Li}= \frac{(N_i^+ - N_i^-)}{\sqrt{2}},
\end{eqnarray}
Thus, the interaction Lagrangian in Eqn.(\ref{Eq:yuk-MD}) can be  written in the new  basis $N_i^\pm$ as
{\small{
\begin{eqnarray}
{\cal W}_D && =  \alpha_D {L}_{e_L} {H_u} \left[\bm{Y} \left(\frac{(N_i^+ + N_i^-)}{\sqrt{2}}\right)\right]_{1} + \beta_D   {L}_{\mu_L} {H_u} \left[\bm{Y} \left(\frac{(N_i^+ + N_i^-)}{\sqrt{2}}\right)\right]_{1^{\prime}} \nn \\
&& + \gamma_D   {L}_{\tau_L} {H_u} \left[\bm{Y} \left(\frac{(N_i^+ + N_i^-)}{\sqrt{2}}\right)\right]_{1^{\prime \prime}} .
\end{eqnarray}}}
Analogously, the pseudo-Dirac interaction term Eqn. (\ref{Eq:yuk-LS}) becomes
{\small{
\begin{eqnarray}
{\cal W}_{LS} &&=\alpha'_D   {L}_{e_L} {H_u} \left[\bm{Y} \left(\frac{(N_i^+ - N_i^-)}{\sqrt{2}}\right)\right]_{1} \frac{\rho^3}{\Lambda^3}+ \beta'_D   {L}_{\mu_L} {H_u} \left[\bm{Y} \left(\frac{(N_i^+ - N_i^-)}{\sqrt{2}}\right)\right]_{1^{\prime}}\frac{\rho^3}{\Lambda^3} \nn \\
 && +\gamma'_D   {L}_{\tau_L} {H_u} \left[\bm{Y} \left(\frac{(N_i^+ - N_i^-)}{\sqrt{2}}\right)\right]_{1^{\prime \prime}} \frac{\rho^3}{\Lambda^3} .
\end{eqnarray}}}
The mass eigenvalues for the new states $N^+$  and $N^-$ can be obtained by diagonalizing the block diagonal form of heavy superfield masses, expressed as
\begin{eqnarray}
M_{RS}\pm \frac{M_R}{2} = \left(\frac{\alpha_{NS} v_\rho}{\sqrt{2}} \pm \frac{\alpha_R v^4_\rho}{4\Lambda^3}\right) \begin{pmatrix}
2y_1 & -y_3 & -y_2 \\ 
-y_3 & 2y_2 & -y_1 \\ 
-y_2 & -y_1 & 2y_3 \\ 
\end{pmatrix}. \label{HMrs} 
\end{eqnarray}
In the above, the anti-symmetric part in $M_{RS}$ is neglected because $\beta_{NS}$ is small compared with $\alpha_{NS}$. The above matrix can be diagonalised through $(M^{\pm})_{\rm diag}=U_{\rm TBM} U_R \left(M_{RS}\pm \frac{M_{R}}{2}\right) U^T_R U^T_{\rm TBM}$, with mass eigenvalues
\begin{eqnarray}
&& M^{\pm}_1 \approx \frac{1}{6}\left(\frac{\alpha_{NS} v_\rho}{\sqrt{2}} \pm \frac{\alpha_R v^4_\rho}{4\Lambda^3}\right)\left(y_1 + 2y_2-\sqrt{9y^2_1 + 12y_1y_2+12 y^2_2}\right),\nonumber\\
&& M^{\pm}_2 \approx \frac{1}{6}\left(\frac{\alpha_{NS} v_\rho}{\sqrt{2}} \pm \frac{\alpha_R v^4_\rho}{4\Lambda^3}\right)\left(y_1 + 2y_2+\sqrt{9y^2_1 + 12y_1y_2+12 y^2_2}\right),\nonumber\\
&& M^{\pm}_3 \approx \frac{1}{3}\left(\frac{\alpha_{NS} v_\rho}{\sqrt{2}} \pm \frac{\alpha_R v^4_\rho}{4\Lambda^3}\right) \left(y_1 + 2y_2 \right).\label{Mrs_masses}
\end{eqnarray}
Here, $U_{\rm TBM}$ is the tribimaximal mixing matrix \cite{Harrison:2002er, Harrison:2002kp} and 
\begin{eqnarray}
U_R \approx \begin{pmatrix}
B_- & \frac{1}{\sqrt{X_-}} & 0\\
0 & 0 & 1\\
B_+ & \frac{1}{\sqrt{X_+}} & 0
\end{pmatrix},
\end{eqnarray}
with
\begin{eqnarray}
B_\pm =-\frac{y_1 + 2y_2 \pm \sqrt{9y^2_1 -12 y_1 y_2 +12 y^2_2}}{2\sqrt{2} (y_1 -y_2)}, ~~{\rm and}~~ X_\pm = \sqrt{1+ B^2_\pm}\;.
\end{eqnarray}
As noticed from Eqn. \eqref{Mrs_masses}, we get three sets of nearly degenerate mass states after diagonalization. We further assume that the lightest pair with TeV scale masses dominantly contribute to the CP asymmetry\footnote{We also have heavier fermions i.e., $N_2^\pm$ and $N_3^\pm$, whose decays can also generate lepton asymmetry. But these heavy fermions decouple early and moreover the asymmetry can be washed out from the inverse decays of lighter fermion mass eigenstates i.e., $\ell H \to N_1^\pm$. Even though we consider the asymmetry generated from other fermions (i.e., $N_2^\pm, N_3^\pm$), the final asymmetry hardly changes upto a maximum of 3 times the asymmetry generated from $N_1^\pm$ in one flavor approximation, which does not really make any appreciable difference in the final result.}. The small mass splitting between the lightest states implies the contribution from one loop self energy of heavy particle decay dominates over the vertex diagram.
The expression for CP asymmetry is given by \cite{Pilaftsis:1997jf,Gu:2010xc}
\begin{eqnarray}
\epsilon_{N^-_i} 
\approx \frac{1}{32\pi^2 A_{N^-_i}}{\rm Im}\left[ \left(\frac{\tilde{M_D}}{v_u}-\frac{\tilde{M}_{LS}}{v_u}\right)^\dagger \left(\frac{\tilde{M}_D}{v_u}+\frac{\tilde{M}_{LS}}{v_u} \right)^2  \left(\frac{\tilde{M}_D}{v_u}-\frac{\tilde{M}_{LS}}{v_u}\right)^\dagger \right]_{ii} \frac{r_N}{r^2_N + 4 A^2_{N^-_i}}.\nn\\
\end{eqnarray}
Here, $\tilde{M}_D= M_D U_{\rm TBM} U_R$,  $\tilde{M}_{LS}= M_{LS} U_{\rm TBM} U_R$ and $\Delta M =M^+_i - M^-_i \approx M_R$. The parameters $r_N$ and $A_{N^-}$ are expressed as 
\begin{eqnarray}
&&  r_N =\frac{{(M^+_i)}^2 - {(M^-_i)}^2}{M^+_i M^-_i} = \frac{\Delta M (M^+_i + M^-_i)}{M^+_i M^-_i}\; ,\nonumber \\
&& A_{N^-} \approx \frac{1}{16\pi}\left[\left(\frac{\tilde{M}_D}{v_u}-\frac{\tilde{M}_{LS}}{v_u}\right)\left(\frac{\tilde{M}_D}{v_u}+\frac{\tilde{M}_{LS}}{v_u}\right) \right]_{ii}.
\end{eqnarray}
It should be noted that because of the imposition of modular symmetry, which plays the role of eliminating the usage of extra  flavon fields,  the CP asymmetry parameter crucially depends on the Yukawa couplings $\bf Y$ = $(y_1,y_2,y_3)$, apart from other free parameters of the model and the flavon VEV $v_\rho$. However, essentially
 there is no freedom in the choice of how much can be the numerical values of the Yukawa couplings as they depend on the real and imaginary part of the modulus $\tau$, which are constrained by the neutrino oscillation data.
In the left (middle) panel of Fig. \ref{CP_var}, we show  the variation of CP asymmetry with the magnitude (argument) of the Yukawa coupling $y_1$ and  right panel projects its behavior with $r_N$. It should be noted that, the CP symmetry in the context of the present model  is broken by the vacuum expectation value of the modulus $\tau$. As  this vacuum expectation value  is related to the CP phases in the PMNS matrix  and the CP asymmetry of leptogenesis, it is generally anticipated that there should be a non-trivial correlation between these observables.  In the bottom panel of  Fig. \ref{CP_var}, we show the correlation plot between the Dirac CP violating phase $\delta_{CP}$   and the CP asymmetry  of  leptogenesis, which depicts no  appreciable correlation between these observables. 

In Table. \ref{tab:leptobench}, we provide benchmark values that satisfy both neutrino mass and required CP asymmetry for leptogenesis \cite{Davidson:2008bu,Buchmuller:2004nz} (to be discussed in the next subsection). 

\begin{center} 
\begin{table}[t!]
\centering

\begin{tabular}{|c|c|c|c|c|c|}  \hline
{$\epsilon^{e}_{N^-}$}  & ~{ $\epsilon^{\mu}_{N^-}$}~& ~$\epsilon^{\tau}_{N^-}$~ & $\epsilon_{N^-}$ & $\Delta M$~(GeV) \\\hline 
 $-9\times 10^{-5}$ & $-2.13 \times 10^{-4}$ & $-2.42 \times 10^{-4}$ &    $-5.45 \times 10^{-4}$ & $2.94\times 10^{-5}$  \\\hline
\end{tabular}

\caption{CP asymmetries and mass splitting obtained from the allowed range of model parameters which satisfy neutrino oscillation data.}
\label{tab:leptobench}
\end{table}
\end{center}
\subsection{One flavor approximation} 
The evolution of lepton asymmetry can be deduced from the dynamics of relevant Boltzmann equations. Sakharov criteria \cite{Sakharov:1967dj} demand the decay of parent fermion to be out of equilibrium to generate the lepton asymmetry. To impose this condition, one has to compare the Hubble rate with the decay rate as follows. 
\begin{equation}
K = \frac{\Gamma_{N^-_1}}{H(T=M^-_1)}\;.
\end{equation}
Here, $H = \frac{1.67 \sqrt{g_\star}~ T^2 }{M_{\rm Pl}}$, with $g_{\star} = 106.75$, $M_{\rm Pl} = 1.22 \times 10^{19}$ GeV. We consider the coupling strength $\left( \approx \left(\frac{\sqrt{2}M_D}{v} U_{\rm TBM}U_R\right)_{ij}\right)$ roughly around $10^{-6}$, where the minimum order of coupling parameters are taken from the numerical analysis section, consistent with neutrino oscillation data. The Boltzmann equations for the evolution of the number densities of right-handed superfield and lepton, written in terms of yield parameter (ratio of number density to entropy density) are given by \cite{Plumacher:1996kc, Giudice:2003jh,Buchmuller:2004nz, Strumia:2006qk, Iso:2010mv}
\begin{eqnarray}
&& \frac{d Y_{N^-}}{dz}=-\frac{z}{s H(M_1^{-})} \left[\left( \frac{Y_{N^-}}{{Y^{\rm eq}_{N^-}}}-1\right)\gamma_D +\left( \left(\frac{{Y_{N^-}}}{{Y^{\rm eq}_{N^-}}}\right)^2-1\right)\gamma_S \right],\nn\\
&& \frac{d Y_{B-L}}{d z}= -\frac{z}{s H(M_{1}^{-})} \left[ \epsilon_{N^-} \left( \frac{Y_{N^-}}{{Y^{\rm eq}_{N^-}}}-1\right)\gamma_D -\frac{Y_{B-L}}{{Y^{\rm eq}_{\ell}}}\frac{\gamma_{D}}{2} \right],
\label{Boltz}
\end{eqnarray}
where $s$ denotes the entropy density, $z = M^-_1/T$ and the equilibrium number densities are given by \cite{Davidson:2008bu}
\begin{eqnarray}
Y^{\rm eq}_{N^-}= \frac{45  g_{N^-}}{4 {\pi}^4 g_\star} z^2 K_2(z), \hspace{3mm} {Y^{\rm eq}_\ell}= \frac{3}{4} \frac{45 \zeta(3) g_\ell}{2 {\pi}^4 g_{\star}}\,.
\end{eqnarray}
Here, $K_{1,2}$ denote modified Bessel functions, $g_\ell=2$ and $g_{N^-}=2$ denote the degrees of freedom of lepton and right-handed superfields respectively. The decay rate $\gamma_D$ is given by
\begin{equation}
\gamma_D = s Y^{\rm eq}_{N^-}\Gamma_D,
\end{equation}
where, $\Gamma_{D} = \Gamma_{N^-} \frac{K_1(z)}{K_2(z)}$. $\gamma_S$ denotes the scattering rate of the decaying particle  i.e.,$N_1^-N_1^- \to \rho \rho$ \cite{Iso:2010mv}\footnote{\begin{equation}
\gamma(ab\leftrightarrow cd)  = \frac{T}{64 \pi^4} \int_{s_{\rm min}}^{\infty} ds ~ \hat{\sigma}(s^\prime) \sqrt{s^\prime} K_1 \left(\frac{\sqrt{s^\prime}}{T}\right)\nn,
\end{equation}
where, $s_{\rm min} = {\rm Max}[(m_a+m_b)^2,(m_c+m_d)^2]$ and $\hat{\sigma}(s^\prime)$ is the reduced cross section with $s^\prime$ denoting the center of mass energy.}. The Boltzmann equation for $Y_{B-L}$ is free from the subtlety of asymmetry getting produced even when $N_1^-$ is in thermal equilibrium i.e., by subtracting the on-shell $N_1^-$ exchange contribution ($\frac{\gamma_{D}}{4}$) from the $\Delta L =2$ process \cite{Giudice:2003jh}.
\begin{figure}[t!]
\begin{center}
\includegraphics[width=0.45\linewidth]{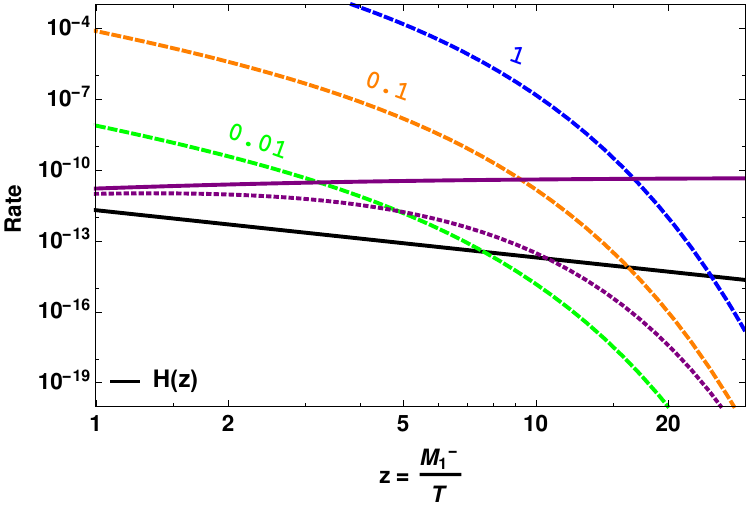}
\hspace{0.5 cm}
\includegraphics[width=0.45\linewidth]{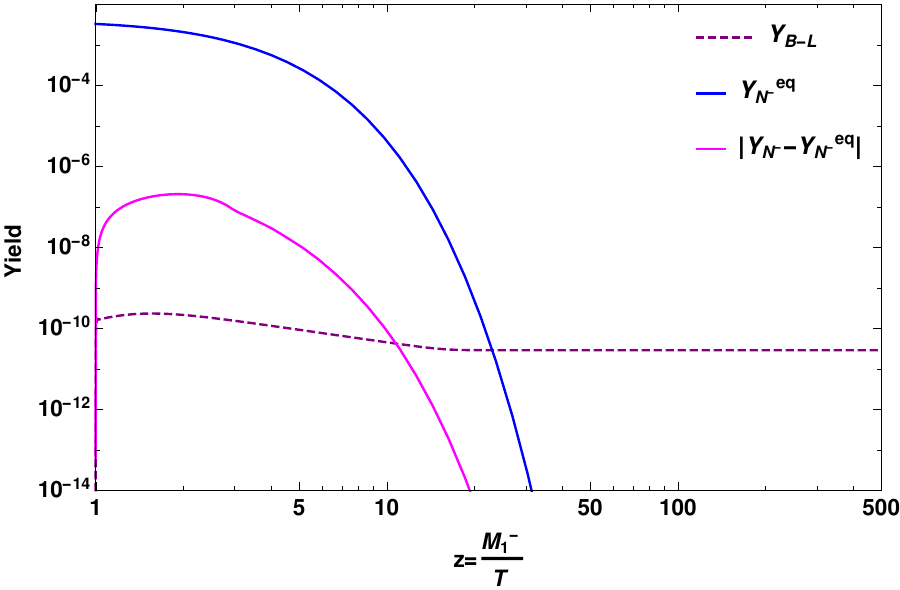}
\caption{Left panel projects the comparison of interaction rates with Hubble expansion, where purple lines correspond to decay (solid), inverse decay (dotted) and scattering rates plotted for various values of Majorana coupling (green, orange, blue). Right panel projects the evolution of $Y_{B-L}$ (dashed) as a function of $z = M_{1}^{-}/{T}$.}\label{yield}
\end{center}
\end{figure}

The interaction rates are compared with Hubble expansion in the left panel of Fig. \ref{yield}. The decay ($\Gamma_D$) and inverse decay $\left(\Gamma_{D}\frac{Y^{\rm eq}_{N^-}}{Y^{\rm eq}_\ell}\right)$ rates are plotted in purple with the coupling strength $\sim 10^{-6}$. The scattering rate $\left(\frac{\gamma_S}{s Y^{\rm eq}_{N^-}}\right)$ for $N_1^-N_1^- \to \rho\rho$ is projected for various set of values for coupling (of Eq. \ref{Eq:yuk-M}), consistent with neutrino oscillation study. For larger Majorana coupling, the scattering process makes $N_1^-$ to stay longer in thermal soup and hence, number density of $N_1^-$ depletes in annihilation rather than decay, generating lesser lepton asymmetry. In one-flavor approximation, the solution of Boltzmann eqns (\ref{Boltz})  using the benchmark given in Table. \ref{tab:leptobench} is projected in the right panel of Fig. \ref{yield} with the inclusion of decay and scattering rates. Once the out-of-equilibrium criteria is satisfied, the decay proceeds slow (over abundance), $Y_{N^-}$ does not trace $Y^{\rm eq}_{N^-}$ (magenta curve)  and the lepton asymmetry (dashed curve) is generated.  The obtained lepton asymmetry gets converted to the observed baryon asymmetry through sphaleron transition, given by \cite{Harvey:1990qw}
\begin{equation}
Y_B = \left(\frac{8N_f + 4 N_H}{22 N_f + 13 N_H}\right)Y_{B-L}.
\end{equation}
Here, $N_f$ denotes the number of superfields generations and $N_H$ is the number of Higgs doublets.
The observed baryon asymmetry is quantified in terms of baryon to photon ratio \cite{Aghanim:2018eyx}
\begin{equation}
\eta = \frac{\eta_b - \eta_{\bar{b}}}{\eta_\gamma} = 6.08 \times 10^{-10}.
\end{equation}
Based on the relation $Y_{B} = (7.04)^{-1} \eta$, the current bound on baryon asymmetry is $Y_{B} \sim 0.86\times 10^{-10}$. 

 We observe the same Yukawas i.e. \textbf{Y}=($y_1, y_2, y_3$) are involved in both Dirac as well as Majorana masses and hence, appear not only in the neutrino phenomenology but also in computation related to leptogenesis. But the values of these couplings are strongly constrained from the real and imaginary part of the complex modulus $\tau$. Thus, the free parameters play an important role in adjusting the parameter space to generate a successful leptogenesis. 
\vspace{0.5 cm}
\subsection{Flavor consideration}

One flavor approximation is probable at high scale ($T>10^{12}$ GeV), where all the Yukawa interactions are out of equilibrium. But for temperatures below $10^{12}$ GeV, various charged lepton Yukawa couplings come into equilibrium and hence flavor effects play a crucial role in generating the final lepton asymmetry. For temperatures below $10^5$ GeV, all the Yukawa interactions are in equilibrium and the asymmetry is stored in the individual lepton sector. The detailed investigation of flavor effects in type-I leptogenesis can be found in the literature \cite{Pascoli:2006ci,Antusch:2006cw,Nardi:2006fx,Abada:2006ea,Granelli:2020ysj,Dev:2017trv}. 

The Boltzmann equation for generating the lepton asymmetry in each flavor is \cite{Antusch:2006cw}
\begin{eqnarray}
\frac{d Y^{\alpha}_{ B-L_\alpha}}{d z}= -\frac{z}{s H(M_1^-)} \left[  \epsilon^\alpha_{N^-} \left( \frac{Y_{N^-}}{{Y^{eq}_{N^-}}}-1\right)\gamma_D-\left(\frac{\gamma^{\alpha}_D}{2}\right)\frac{A_{\alpha \alpha}Y^\alpha_{\rm B-L_\alpha}}{{Y^{eq}_{\ell}}}\right],
\end{eqnarray}
where, $\epsilon^\alpha_{N^-}$ represents the CP asymmetry in each lepton flavor and 
\begin{equation}
\gamma_D^\alpha = s Y_{N^-}^{eq}\Gamma_{N^-}^\alpha \frac{K_1(z)}{K_2(z)}, \quad \gamma_D = \sum_\alpha \gamma^\alpha_D,\nn\\
\end{equation}
The matrix $A$ is given by \cite{Nardi:2006fx}, 
\begin{equation}
A=\begin{pmatrix}
-\frac{221}{711} && \frac{16}{711} && \frac{16}{711}\\
\frac{16}{711} && -\frac{221}{711} && \frac{16}{711}\\
\frac{16}{711} && \frac{16}{711}  && -\frac{221}{711} \\
\end{pmatrix}.\nn\\
\end{equation}
\begin{figure}[t!]
\begin{center}
\includegraphics[width=0.45\linewidth]{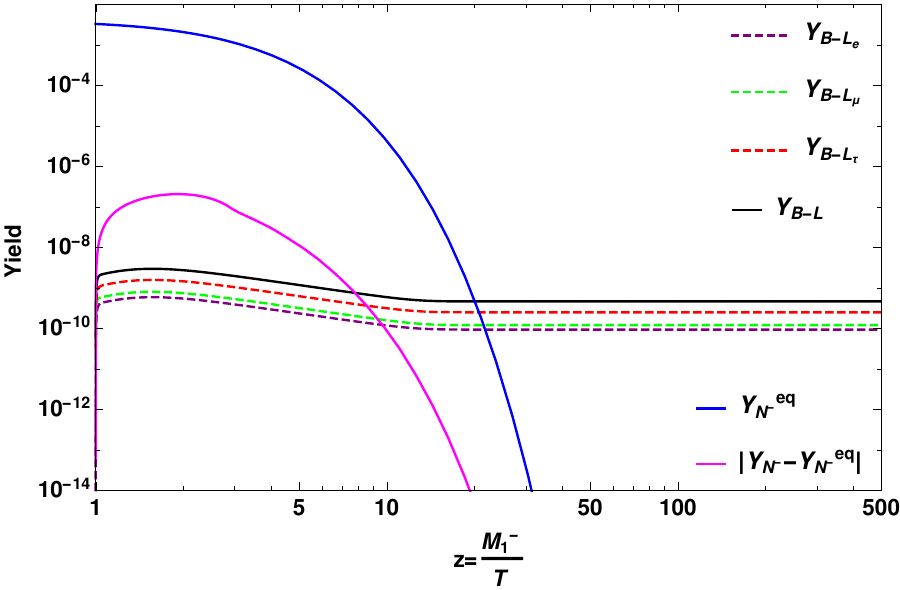}
\hspace{0.5 cm}
\includegraphics[width=0.45\linewidth]{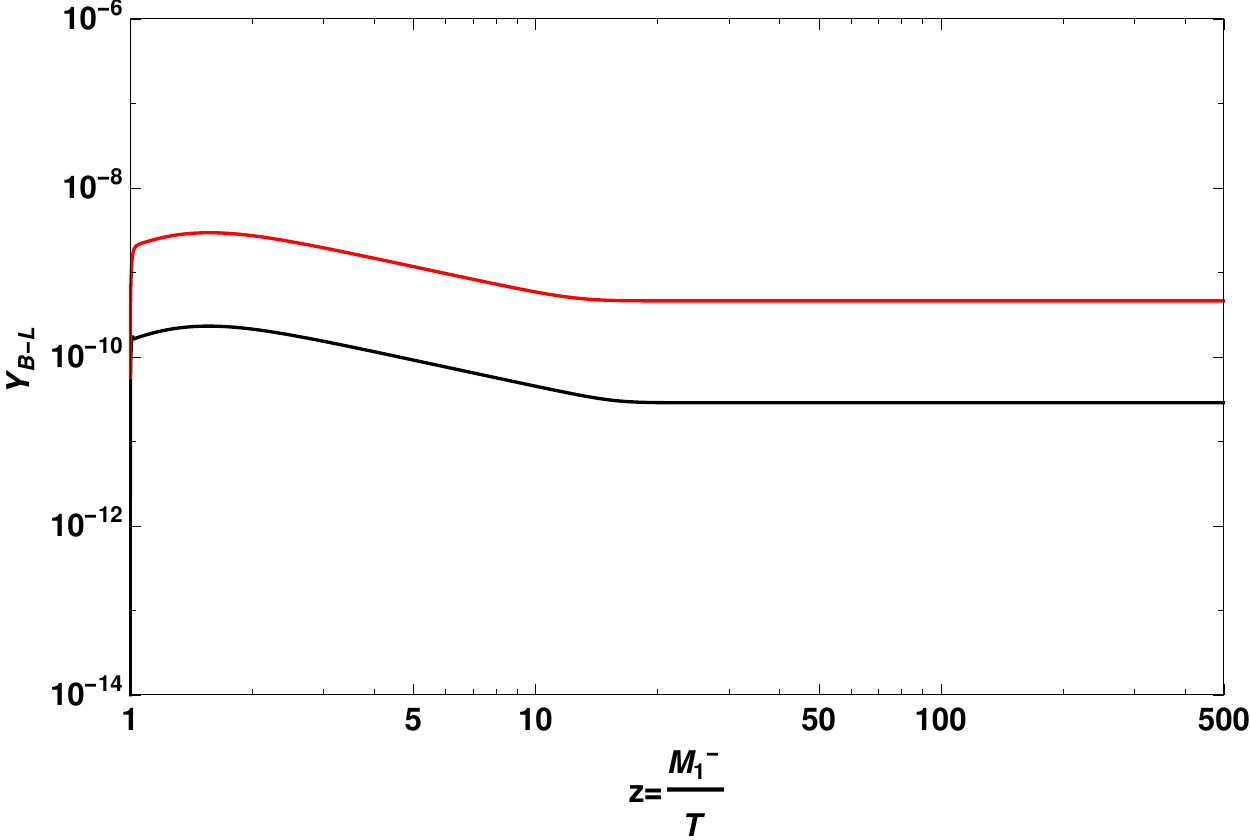}
\caption{The left panel displays yield with inclusion of flavor effects. The right panel shows the enhancement in the yield due to three-flavor calculation (red curve) over one-flavor approximation (black curve).}
\label{yield2}
\end{center}
\end{figure}
From the benchmark shown in Table. \ref{tab:leptobench}, we project the $B-L$ yield with flavor consideration in the left panel of Fig. \ref{yield2}. It is clear that a notable enhancement in $B-L$ asymmetry is obtained in case of flavor consideration (red curve) over one flavor approximation (black curve), as displayed in the right panel. This is because, in one flavor approximation the decay of heavy fermion to a specific lepton flavor final state can get washed out by the inverse decays of any flavor unlike the flavored case \cite{Abada:2006ea}.

\vspace*{0.2 true in}
\section{Comment on collider studies}
\label{collider}
Here, we briefly comment on the most promising collider signature of heavy pseudo-Dirac neutrinos without going into any detailed estimation, in the context of the present model.  In the linear seesaw scenario the $M_{LS}$ is the lepton number violating term \cite{han2021full} therefore its mass scale is naturally small. Also the effective Majorana neutrino mass matrix as shown in eqn.(\ref{mass}) for active neutrino where the smallness of $m_\nu$ is attributed due to $M_{LS}$ being the pseudo-Dirac neutrino mass term and  further suppressed by the ratio of $M_{D}$ and $M_{RS}$. Hence, the seesaw scale can be lowered to TeV range which is experimentally accessible at LHC. The trilepton plus missing energy process as mentioned in eqn.(\ref{tri}), which can be studied at colliders, is an interesting mechanism involving heavy pseudo-Dirac neutrinos \cite{Das:2014jxa}:
\begin{equation}
\sigma (pp \to \ell^\pm \ell^\pm \ell^\mp+ \slashed{E}_T) \ = \sigma(pp \to W^{*\pm} \to N \ell^\pm) \times \mathrm{Br}(N \to \ell^\mp  W^{\pm})\times \mathrm{Br}(W^{\pm} \to \ell^\pm   + \nu(\bar \nu)),
\label{tri}
\end{equation} 
where it is assumed that the heavy neutrinos are heavier than the $W$ boson, so that 
the two-body decay process $N \to \ell  W$ is kinematically allowed, followed by the on-shell $W$ decaying into SM leptons. 
Its viability is essentially determined by firstly, large mixing between active-sterile neutrinos i.e. $\mathsf{\Theta}_{\nu RS} \simeq \sqrt{\nicefrac{m_\nu}{M_{RS}}} \leq 10^{-6}$ \cite{Aguilar-Saavedra:2009fxa}, secondly, masses of heavy pseudo-Dirac neutrinos ranging from few [GeV-TeV], and finally its production mechanism.

\section{Conclusion}
\label{sec:con}
We have investigated a modular form of $A_4$ flavor symmetry that reduces the complications of accommodating multiple flavons. The present model includes three right-handed and three left-handed heavy superfields to explore the neutrino phenomenology within the framework of linear seesaw in SUSY context.
We have considered the Yukawa couplings to transform non-trivially under modular $A_4$ group, which replaces the role of conventional flavon fields. This leads to a specific flavor structure of the neutrino mass matrix and helps in studying the neutrino mixing. We numerically diagonalized the neutrino mass matrix to obtain an allowed region for the model parameters compatible with the current $3\sigma$ limit of oscillation data. The flavor structure of heavy superfields gives rise to six doubly degenerate mass eigenstates and hence, to explain leptogenesis, we introduced a higher dimensional mass term for the right-handed superfields to generate a small mass splitting. We obtained a non-zero CP asymmetry from the decay of lightest heavy fermion eigen state and the self energy contribution is partially enhanced due to the small mass difference between the two lighter heavy fermion superfields. Using a specific benchmark of model parameters consistent with oscillation data, we solved  coupled Boltzmann equations to obtain the evolution of lepton asymmetry at TeV scale which comes out to be of the order $\approx 10^{-10}$, which is sufficient to explain the present baryon asymmetry of the Universe. Furthermore, we have also discussed the enhancement in asymmetry with flavor consideration.  The promising collider signature of the   heavy pseudo-Dirac neutrinos is the trilepton plus missing energy, which depend crucially on the mixing between the light active and pseudo-Dirac neutrinos, mass of these heavy neutrinos and their production mechanism. 

\acknowledgments

MKB and SM want  to acknowledge DST for its financial help. RM and SS would like to that University of Hyderabad for financial support through IoE project grant No.  UoH/IoE/RC1/RC1-20-012. RM  acknowledges the support from  SERB, Government of India, through grant No. EMR/2017/001448.  The computational work done at CMSD, University of Hyderabad is duly acknowledged.


\section*{Appendix}

 $\bar\Gamma$ is the modular group which attains a  linear fractional transformation
$\gamma$ which acts on modulus  $\tau$ 
linked to the upper-half complex plane whose transformation is given by
\begin{equation}\label{eq:tau-SL2Z}
\tau \longrightarrow \gamma\tau= \frac{a\tau + b}{c \tau + d}\ ,~~
{\rm where}~~ a,b,c,d \in \mathbb{Z}~~ {\rm and }~~ ad-bc=1, 
~~ {\rm Im} [\tau]>0 ~ ,
\end{equation}
where it is isomorphic to the transformation$PSL(2,\mathbb{Z})=SL(2,\mathbb{Z})/\{I,-I\}$.
The $S$ and $T$ transformation helps in generating the modular transformation defined by
\begin{eqnarray}
S:\tau \longrightarrow -\frac{1}{\tau}\ , \qquad\qquad
T:\tau \longrightarrow \tau + 1\ ,
\end{eqnarray}
and hence the algebric relations so satisfied are as follows,
\begin{equation}
S^2 =\mathbb{I}\ , \qquad (ST)^3 =\mathbb{I}\ .
\end{equation}

Here, series of groups are introduced, $\Gamma(N)~ (N=1,2,3,\dots)$ and defined as
 \begin{align}
 \begin{aligned}
 \Gamma(N)= \left \{ 
 \begin{pmatrix}
 a & b  \\
 c & d  
 \end{pmatrix} \in SL(2,\mathbb{Z})~ ,
 ~~
 \begin{pmatrix}
  a & b  \\
 c & d  
 \end{pmatrix} =
  \begin{pmatrix}
  1 & 0  \\
  0 & 1  
  \end{pmatrix} ~~({\rm mod} N) \right \}
 \end{aligned} .
 \end{align}
Definition of $\bar\Gamma(2)\equiv \Gamma(2)/\{I,-I\}$ for $N=2$.
Since $-I$ is not associated with $\Gamma(N)$
  for $N>2$ case, one can have $\bar\Gamma(N)= \Gamma(N)$,
  which are infinite normal subgroup of $\bar \Gamma$ known as principal congruence subgroups.
   Quotient groups come from the finite modular group defined as
   $\Gamma_N\equiv \bar \Gamma/\bar \Gamma(N)$.
Imposition of $T^N=\mathbb{I}$, is done for these finite groups $\Gamma_N$.
 Thus, the groups $\Gamma_N$ ($N=2,3,4,5$) are isomorphic to
$S_3$, $A_4$, $S_4$ and $A_5$, respectively \cite{deAdelhartToorop:2011re}.
$N$ level modular forms  are 
holomorphic functions $f(\tau)$ which are transformed under the influence of $\Gamma(N)$ as follows:
\begin{equation}
f(\gamma\tau)= (c\tau+d)^k f(\tau)~, ~~ \gamma \in \Gamma(N)~ ,
\end{equation}
where $k$ is the modular weight.

Here the discussion is all about the modular symmetric theory. This paper comprises of  $A_4$ ($N=3$) modular group. 
A field $\phi^{(I)}$ transforms under the modular transformation of Eq.(\ref{eq:tau-SL2Z}),  as
\begin{equation}
\phi^{(I)} \to (c\tau+d)^{-k_I}\rho^{(I)}(\gamma)\phi^{(I)},
\end{equation}
where  $-k_I$ represents the modular weight and $\rho^{(I)}(\gamma)$ signifies an unitary representation matrix of $\gamma\in\Gamma(2)$.

The scalar fields$'$ kinetic term is as follows
\begin{equation}
\sum_I\frac{|\partial_\mu\phi^{(I)}|^2}{(-i\tau+i\bar{\tau})^{k_I}} ~,
\label{kinetic}
\end{equation}
which doesn't change under the modular transformation and eventually the overall factor is absorbed by the field redefinition.
Thus, the Lagrangian should be invariant under the modular symmetry.

The modular forms of the Yukawa coupling {\bf Y} = {$(y_{1},y_{2},y_{3})$}  with weight 2,   which transforms
as a triplet of $A_4$ can be expressed in terms of Dedekind eta-function  $\eta(\tau)$ and its derivative \cite{Feruglio:2017spp}:
\begin{eqnarray} 
\label{eq:Y-A4}
y_{1}(\tau) &=& \frac{i}{2\pi}\left( \frac{\eta'(\tau/3)}{\eta(\tau/3)}  +\frac{\eta'((\tau +1)/3)}{\eta((\tau+1)/3)}  
+\frac{\eta'((\tau +2)/3)}{\eta((\tau+2)/3)} - \frac{27\eta'(3\tau)}{\eta(3\tau)}  \right), \nonumber \\
y_{2}(\tau) &=& \frac{-i}{\pi}\left( \frac{\eta'(\tau/3)}{\eta(\tau/3)}  +\omega^2\frac{\eta'((\tau +1)/3)}{\eta((\tau+1)/3)}  
+\omega \frac{\eta'((\tau +2)/3)}{\eta((\tau+2)/3)}  \right) , \label{eq:Yi} \\ 
y_{3}(\tau) &=& \frac{-i}{\pi}\left( \frac{\eta'(\tau/3)}{\eta(\tau/3)}  +\omega\frac{\eta'((\tau +1)/3)}{\eta((\tau+1)/3)}  
+\omega^2 \frac{\eta'((\tau +2)/3)}{\eta((\tau+2)/3)}  \right)\,.
\nonumber
\end{eqnarray}
%
It is interesting to note that the couplings those are defined as singlet under $A_4$ start from $-k=4$ while they are zero if $-k=2$.

\bibliographystyle{my-JHEP}
\bibliography{linear}

\end{document}